\documentclass{aa}
\usepackage{txfonts}
\usepackage{epsfig}
\usepackage{subfigure}
\usepackage{multirow}
\usepackage{hyperref}
\usepackage{afterpage}

\usepackage{times}
\usepackage{natbib}
\usepackage{url}
\bibpunct{(}{)}{;}{a}{}{,}

\begin{document}

\title{MoCA: A Monte Carlo code for Comptonisation in Astrophysics. I. Description of the code and first results}

\author{Francesco Tamborra\inst{1}, Giorgio Matt\inst{2}, Stefano Bianchi\inst{2}, Michal Dov\v ciak\inst{1}}

\offprints{Francesco Tamborra\\ \email{francesco.tamborra@asu.cas.cz}}

\institute{Astronomical Institute of the Czech Academy of Sciences, Bo\v cn\'i II 1401/1, 14100 Prague, Czech Republic
\and Dipartimento di Fisica e Matematica, Universit\`a degli Studi Roma Tre, via della Vasca Navale 84, 00146 Roma, Italy
}

\date{Received / Accepted }

\authorrunning{F. Tamborra et al.}
\titlerunning{MoCA: a Monte Carlo code for Comptonisation in Astrophysics. }

\abstract 
{We present a new Monte Carlo code for Comptonisation in Astrophysics (MoCA). To our knowledge MoCA is the first code that uses a single photon approach in a full special relativity scenario, and including also Klein-Nishina effects as well as polarisation. In this paper we describe in detail how the code works, and show first results from the case of extended coronae in accreting sources Comptonising the accretion disc thermal emission. We explored both a slab and a spherical geometry, to make comparison with public analytical codes more easy.
Our spectra are in good agreement with those from analytical codes for low/moderate optical depths, but differ
significantly, as expected, for optical depths larger than a few. Klein-Nishina effects become relevant above 100 keV depending on the optical thickness and thermal energy of the corona. We also calculated the polarisation properties for the two
geometries, which show that X-ray polarimetry is a very useful tool to discriminate between them. }

\keywords{XRBs, AGN, Comptonisation, Polarisation}
\maketitle 
\section{Introduction}
Up-scattering of low-energy photons by Inverse Compton on a hot gas of electrons (i.e. Comptonisation) is an important mechanism in astrophysics.
In particular, in accreting sources such as X-Ray Binaries (XRBs) and Active Galactic Nuclei (AGN) this mechanism is believed to be responsible for their hard X-ray emission \citep{Sunyaev1980}: soft photons produced by the accretion disc are Comptonised by a corona of hot electrons.
The typical Comptonisation spectrum of AGN can be approximated well by a power-law with photon energy index $\Gamma \sim 1.5-2.0$ extending from a few to hundreds keV with an exponential cut-off at high energies (e.g. \cite{Perola2002}, \cite{Dunn2010}).
The corona of electrons is characterised by its temperature, $k\,T_e$, its geometry and its density (or optical depth, $\tau$).
Thanks to the NuSTAR \citep{Harrison2013} high quality broad-band (3-79 keV)  spectra, the coronal properties of AGN have been the subject of several studies in the last years (e.g. \cite{Fabian2015}, \cite{Tortosa2018}, and references therein). The measured cut-off energies, even if with a large scatter and in some cases with only lower limits, cluster to values between 100 and 250 keV, in agreement with previous measurements performed by BeppoSAX (e.g \cite{Perola2002}), INTEGRAL and Swift missions \citep{Molina2013}. Using the approximate relation stating that the observed high-energy cut-off correspond to  approximately two to three times the thermal energy of the corona (\cite{Petrucci2000}, \cite{Petrucci2001}) (which has been proved to be accurate for an extended slab geometry but it is often used for coronae of any geometry and size), the extrapolated $k\,T_e$ is of the order of $\sim 50-150$ keV.
Furthermore, if the corona is really compact as evidence, mainly from timing arguments \citep[][and reference therein]{Uttley2014}, suggests, these values would put the coronal parameters very close to the boundary for pair production \citep{Fabian2015}.
XRBs, on the other hands, do not in general have stable X-ray emission and cycle from a state in which the thermal emission of the disc is dominant over a steep power-law tail at high-energy with typical spectral index  $\Gamma \sim 2.5$ (the so-called soft-state) and a state characterised by a strong and flatter power-law emission with a spectral indices and high energy cutoffs similar to those  observed in AGN (i.e. $\Gamma \sim 1.5 -2.0$, \cite{Fabian2015}) but often associated with the presence of a jet observed in the radio which correlates non-linearly with the X-ray luminosity (the so-called hard-state) \citep{Fender2001}.
One of the most popular interpretations, also supported by the study of the variability of these sources in the two states, is that in the hard state the accretion disc is truncated at some radius and the inner flow is filled by the corona which Comptonises the thermal emission producing the strong power-law emission at high energies and a stable jet is also formed.
As the accretion rates increases, the accretion disc extends up to the innermost stable circular orbit (ISCO) and a very small but hot corona produces a much steeper power-law at high energies and the Lorentz factor of the jet rises sharply, before the jet is suppressed in a soft disc-dominated state \citep{Fender2004}.

In the last three decades, the Comptonisation problem has been addressed by many authors in different ways and the results
of these studies are now often available in the form of \texttt{XSPEC}\footnote{\texttt{XSPEC} is a command-driven, interactive, X-ray spectral-fitting programme. More information can be found at: \url{http://heasarc.gsfc.nasa.gov/xanadu/xspec/}} models which are widely used by astronomers to fit X-ray spectra. 
Most, if not all, of these \texttt{XSPEC} models have been obtained by analytically or numerically solving the nonlinear radiative transfer equations under certain approximations and/or simplifying assumptions. 
Different approaches were adopted, from solving the Kompaneets equations in the diffusion regime ($\tau \gg 1$) and in the optically thin regime ($\tau < 1$) and then combining the asymptotic solutions to using iterative procedures to solve the radiative transfer equations for consecutive scattering orders.
The advantage of a semi-analytical approach is its capacity to build a grid of solutions for a large range of coronal parameters in order to fit the data. The comparison made between this approach and the first Monte Carlo simulations \citep{Stern1995} was showing a good agreement in most regimes while offering a much faster method with respect to the computational time needed on the most advanced calculators at that time. However, this approach has several limitations. In particular it is less reliable for specific combinations of coronal parameters and it basically consists in a mono-dimensional description of the radiative process. It is beyond the purposes of this paper to discuss in detail all the semi-analytical methods for solving the Comptonisation problem and we refer the reader to the complete analysis made by \cite{Poutanen1996} who, while presenting their calculations (which led to \texttt{compPS} XSPEC model), also analysed the different approaches and results of their predecessors (e.g. \cite{Titarchuk1995, HuaTitarchuk1995}).

In the context of this scenario we developed \texttt{MoCA}: a Monte Carlo code for Comptonisation in Astrophysics.
Our code differs from other models because instead of solving radiative transfer equations we followed every photon from the source to the observer using the Monte Carlo method.
This approach allows to explore the whole space of parameters characterising the Comptonising medium without any particular limitation.
\texttt{MoCA} includes special relativity and all quantum effects such as the Klein-Nishina differential cross-section and scattering angles distribution and the Maxwell-J\"{u}ttner electron distribution. 
The code is modular and, even if in this paper we discuss only the case of the hot coronae around accreting black holes, it can be easily modified to be applied to different Comptonisation problems in astrophysics.
An important feature of \texttt{MoCA} is polarisation: for every photon we register its Stokes parameters, which allows us to calculate the polarisation degree and polarisation angle. This feature is particularly relevant now that the SMEX NASA mission IXPE \citep{Weisskopf2016} has been approved, as onboard it will have the first new generation X-ray polarimeter after 40 years since the last X-ray polarisation measurements.
Polarimetry combined to spectroscopical analysis has the potential to infer the geometry of the corona which can help to understand its origin. For this work we have considered two geometries for the corona: a spherical corona and a slab corona. In the first case (if it is very compact) one might expect the corona to be the base of a jet (or an aborted jet) while in the latter scenario it can be originated by perturbations in the disc.
This version of the code does not include General Relativity (GR) effects therefore, even if we consider the geometrical effect of a compact corona on the spectrum and polarisation signal, we mostly focus on the case of extended coronae (this allows us to compare with the above mentioned XSPEC models which do not include GR as well and consider only extended coronae). Nonetheless a version of MoCA coupled with a GR ray-tracing code is being developed and will be the subject of a future paper.

MoCa is not the first code to deal with Comptonisation (including polarisation) with a Monte Carlo approach. A similar approach was used a few years ago by Schnittman and collaborators \citep{Schnittman2010}. However, to increase the computational efficiency of the calculations they used photon packets that include the entire broad-band spectrum instead of a single photon. For this reason their code works in Thomson regime and the energy exchange is taken into account only by the boost in the reference frame of the rotating corona and then in the reference frame of the electron while the scattering is always elastic.
More recently, Beheshtipour and collaborators \citep{Beheshtipour2017} expanded upon the work of Krawczynski \citep{Krawczynski2012} and manage to include proper Klein-Nishina treatment for Comptonisation in their code (among other things such as the contribution of non-thermal electrons in the corona and cyclotron photons in addition to thermal photons as seeds) following the same photon packets approach as in Schnittmann.

The paper is organised as follows:
a brief description of the code is given in Section 2, with full details  in the Appendix.
In the following sections we show the results obtained with \texttt{MoCA} when applied to the case of Comptonisation for accreting sources: Section 3 is focussed on spectra while in Section 4 we focus on the polarisation signal.
Lastly, in Section 5, we summarise our work.

\section{The code}
\texttt{MoCA} is written in Fortran 2003 and it is modular, allowing for applications to different systems of astrophysical interest with minor changes.
In this section we give a brief description of how the code works when applied to accreting systems while a more detailed description and all the formulae and test we performed can be found in the Appendix.

We have implemented two different geometries for the source of the seed photons: either a point-like along the symmetry axis of the system, or an accretion disc.
For the point-like source the emission can be monochromatic or a power law or a Wien tail while for the disc it is a multi-temperature black body (MTBB from now on), with the radial dependence of the temperature as for a Shakura-Sunyaev \citep{Shakura1973} geometrically thin and optically thick disc.
The seed photons can be emitted isotropically at a given radius or, as for infinitely thick atmospheres, they can be emitted mostly perpendicularly to the plane of the disc (i.e. limb darkening effect).
When polarisation calculations are 
turned on, a polarisation vector, always perpendicular to the wave vector os the seed photons, is created and it can be randomly orientated (for unpolarised radiation) when the emission is isotropic or polarised accordingly to Chandrasekhar predictions \citep{Chandrasekhar1960} for infinitely thick atmospheres when the emission is limb darkened.

The corona of electrons surrounding the disc and responsible for the Comptonisation of seed photons is characterised by its thermal energy, $k\,T_e$, its geometry and its density (or optical depth).
The first type of geometry we implemented in the code is the spheroid, characterised by the semi-diameter along the spin axis, $H_{cor}$, and the semi-diameter along the azimuthal plane, $R_{cor}$.
\begin{figure}
\includegraphics[width=88.mm,clip=]{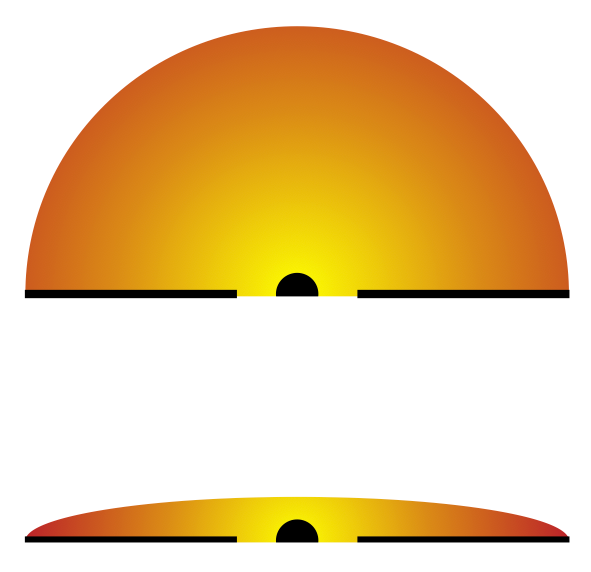}
\caption{\label{geometries} Sketch of the corona geometries implemented in \texttt{MoCA}, compatible with those implemented in most of the analytical Comptonisation models available for \texttt{XSPEC}.}
\end{figure}
As an illustration, we have considered two geometries for the corona: a very oblate spheroid with $H_{cor}/R_{cor} = 0.01$ (to which we refer as a slab) and a spherical one ($H_{cor} = R_{cor}$), both covering the entire disc (see Fig.~\ref{geometries}, not to scale). 
Our simulations considered only one side, assuming the disc being infinite.
We selected these two geometries because they are the geometries implemented in \texttt{XSPEC} models and also because they represent a good choice for testing the polarisation having different symmetry properties.
The velocity of free electrons composing the corona is calculated using the Maxwell-J\"{u}ttner distribution as given in  \cite{Titarchuk1995} which naturally reduces to the classic Maxwell-Boltzmann distribution for non-relativistic electrons.

MoCA treats the scattering with Klein-Nishina corrections which makes the computation more time-demanding mainly because
of the switching between the reference frame of the disc and the reference frame of the electron and also because 
of the complexity of the  distributions of the scattering angles. 
However, this approach allows us to explore the full parameter space without any particular limitation and as we show in the next section that it can affect the shape of the spectrum in particular scenarios.

For each photon which reaches the observer at infinity the Stokes parameters Q and U are calculated and registered together with the direction, energy, and number of scatterings experienced by the photon before escaping. In principle, however, any additional information can be recorded if needed.

\section{The spectrum}
As an illustration, in this section we show some of the results which can be obtained with \texttt{MoCA}. We focus here on the spectral results while in the next chapter we focus on the polarisation signal.

\begin{figure}
\includegraphics[width=88.mm,clip=]{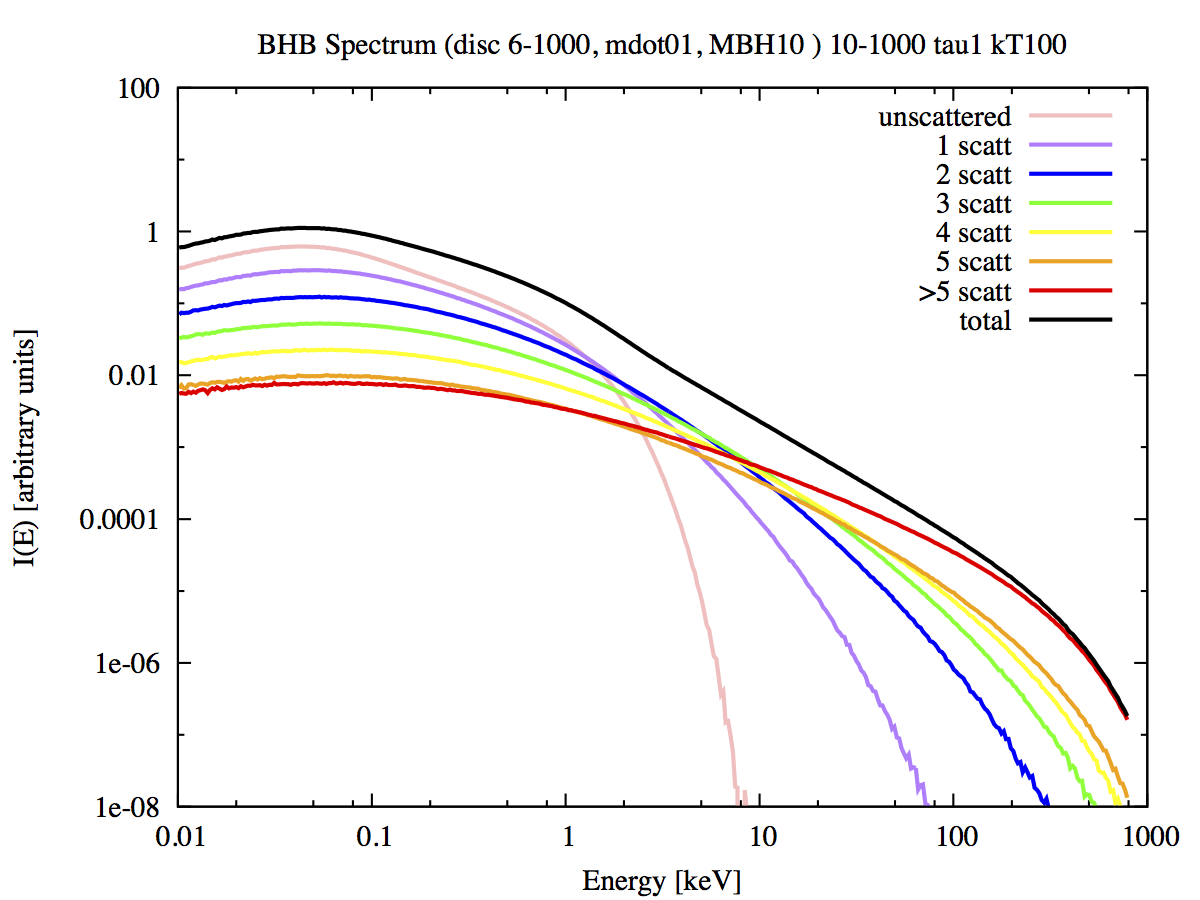}
\caption{\label{spectrum_multi} Spectrum of a $10 \, M_{\odot}$ black hole accreting at $\dot{m} =0.1$ in Eddington luminosity units and Comptonised by a slab corona with  $ \tau = 1$ and $kT_e = 100$ keV. The pink line represents the unscattered component (i.e. the soft X-rays disc emission reaching the observer) while the coloured lines represent the different scattering order components as specified in the figure. The total spectrum observed at infinity (black line) is a power-law with a high energy cut-off.}
\end{figure}

In Fig.~\ref{spectrum_multi} we show the spectrum (photon counts in arbitrary units) produced by Comptonisation of soft disc photons from a slab corona with $ \tau = 1$ and $kT_e = 100$ keV. 
The corona covers the whole disc, $R_{cor}$ = 1000 $r_g$, while on the spin-axis the height is $H_{cor}$ = 10 $r_g$.
We chose such a large value for $R_{cor}$  because we wanted the corona to cover the whole disc (as in XSPEC Comptonisation models).
This value is the one we chose for the disc and it is large enough to be considered infinite (another assumption made by the models we want to compare to) and therefore avoid edge effects at lower energies.
The pink line shows the unscattered component reaching the observer and arising from a disc with inner and outer radius equal to 6 and 1000 $r_g$ respectively, surrounding a $10 M_{\odot}$ black hole accreting at $\dot{m} =0.1$ in Eddington luminosity units. Coloured lines show the different scattering order components forming the total power-law spectrum with a high-energy cut-off (black line). 
The spectrum in Fig.~\ref{spectrum_multi} was obtained integrating along all lines of sight (to increase 
the statistics) so it is not an observable. However from the spectroscopic point of view the inclination does not change the shape of the spectrum significantly.

\begin{figure}
\includegraphics[width=88.mm,clip=]{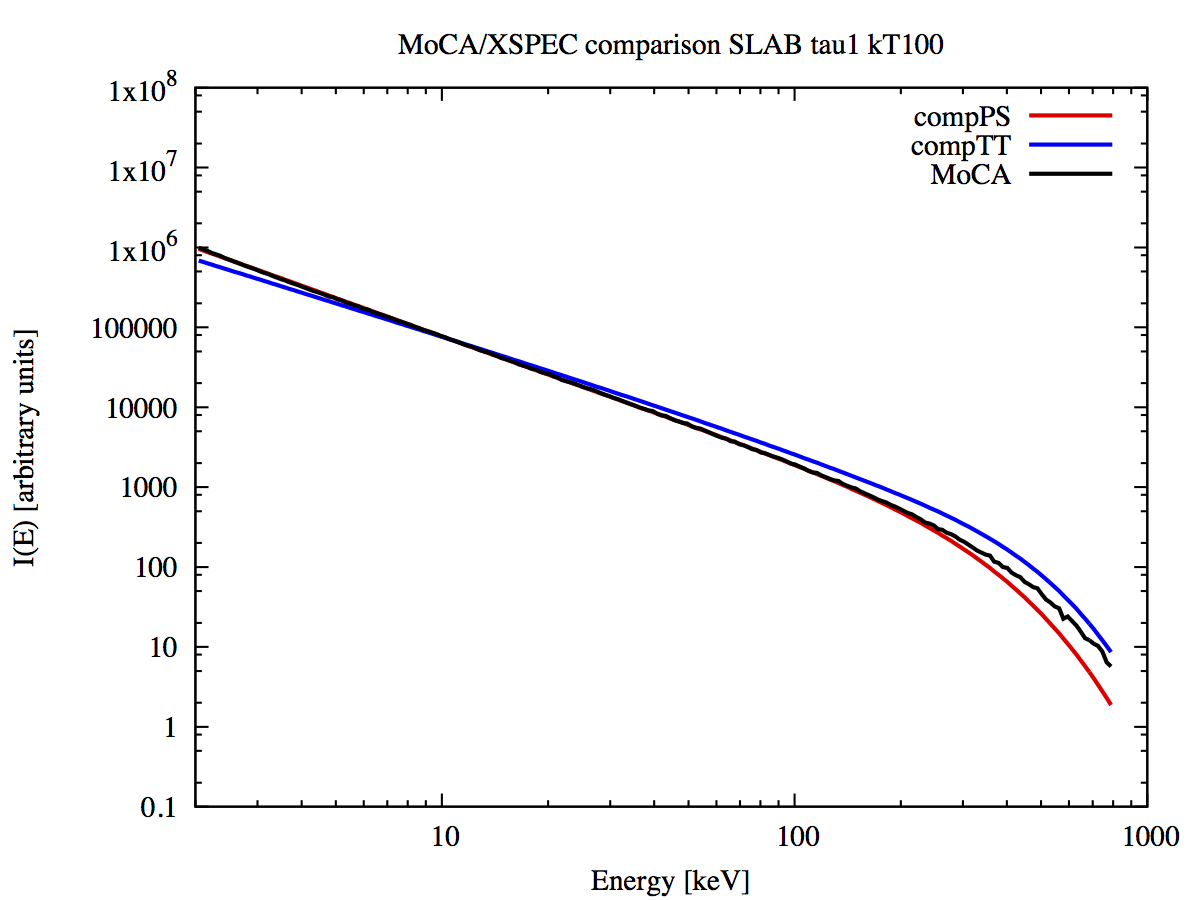}
\caption{\label{xspec_goodcomp} Spectral comparison between \texttt{MoCA} (black line) and the two semi-analytical Comptonisation model \texttt{compPS} and \texttt{compTT} as indicated inside the figure for the case of a slab corona with $\tau = 1$ and $kT_e = 100$ keV observed at an inclination of 60 degrees. See text for more details.}
\end{figure}

\begin{figure}
\includegraphics[width=88.mm,clip=]{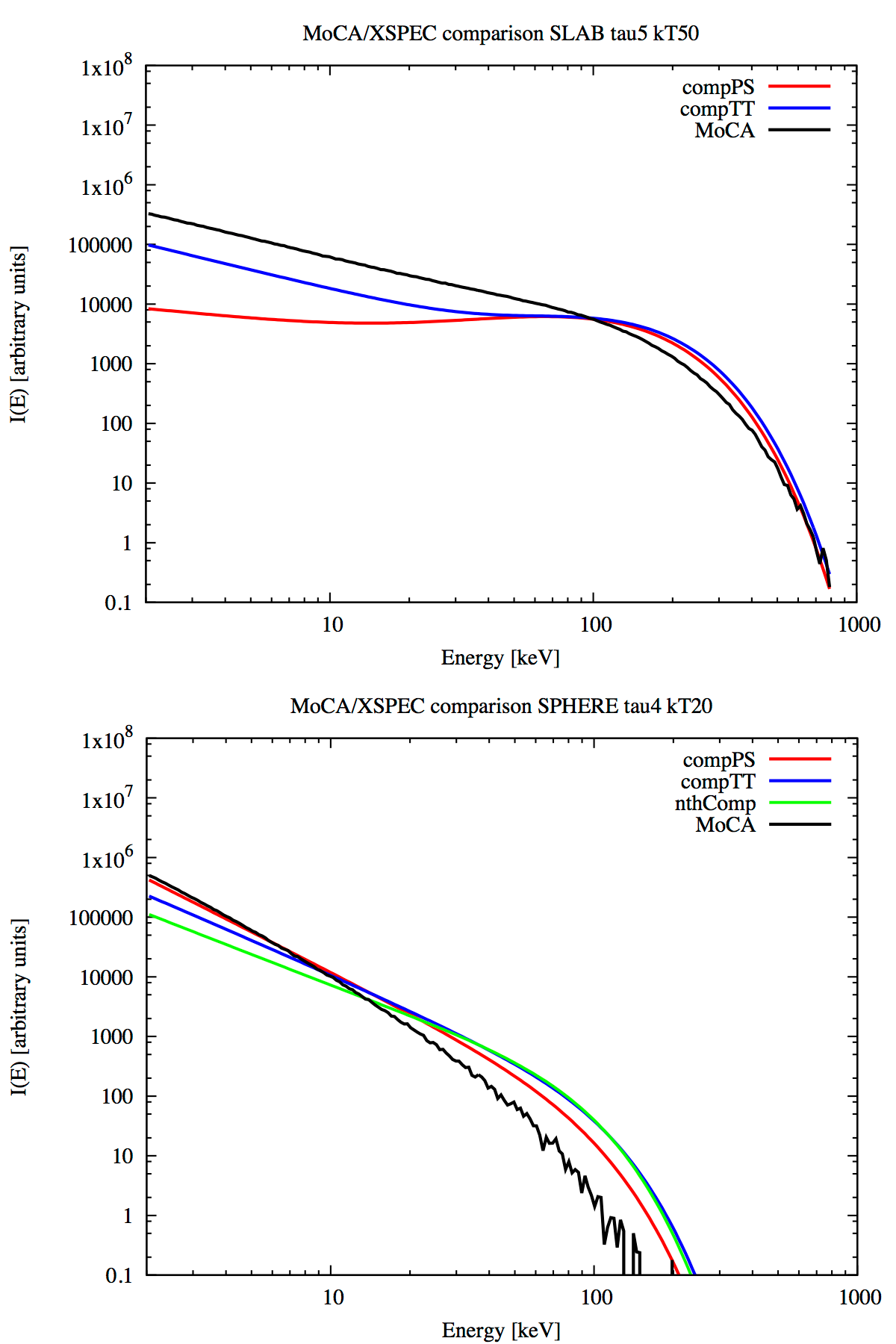}
\caption{\label{xspec_badcomp} Spectral comparison between \texttt{MoCA} (black line) and the semi-analytical Comptonisation model available in \texttt{XSPEC} as indicated inside the figure. \textit{Upper panel} shows the case of a slab corona with $\tau = 5$ and $kT_e = 50$ keV observed at an inclination of 60 degrees while \textit{lower panel} shows the case of a spherical corona with $\tau = 4$ and $kT_e = 40$ keV at the same inclination. In the latter case we performed the comparison with \texttt{Nthcomp} as well.}
\end{figure}

We compared our spectrum with three of the most commonly used Comptonisation models in \texttt{XSPEC}: \texttt{compPS} \citep{Poutanen1996}, \texttt{compTT} \citep{Titarchuk1995, HuaTitarchuk1995} and \texttt{Nthcomp} \citep{Zdziarski1996}. The former two models include the geometry of the corona as one of their parameters, while \texttt{Nthcomp} uses the geometry-independent photon index $\Gamma$.
The comparison has been performed by fitting each model on \texttt{MoCA} `data' between 2 and 800 keV keeping  frozen all the parameters at the value we used in our MoCA simulations (i.e. $\tau$, $kT_e$, and the geometry of the corona) with the exception of the normalisation. 
We decided to cut-off from the fit the thermal component because some models such as \texttt{compTT} use a simple Wien tail while we use the more accurate MTBB profile (as in \texttt{compPS}). Even choosing a value for \texttt{compTT} putting the peak of thermal emission coincident with the one in MoCA ($k\, T_{bb} \sim 0.02 \, keV$ in this particular case) the profiles are quite different and we did not wanted the fit to be driven by differences in the thermal component because we are interested in comparing the slope of the power-law and the high-energy cut-off (i.e. the Comptonisation spectrum).
For \texttt{Nthcomp} we inverted the relation used in the code which links $\tau$ to $\Gamma$ and $kT_e$ \citep[in the Appendix of][]{Zdziarski1996} in order to calculate the spectral index as a function of the other two parameters and we kept it frozen. However, to our understanding, that relation is valid 
in a uniform sphere, so we performed the comparison with \texttt{Nthcomp} only for the spherical corona.
In Fig.~\ref{xspec_goodcomp} we show the comparison between \texttt{MoCA} (black line) and the two Comptonisation models as indicated inside the figure for the case of a slab corona with $\tau = 1$ and $kT_e = 100$ keV. 
In this regime we see a good agreement between \texttt{MoCA}, \texttt{compPS} and \texttt{compTT}: all of them reproduce the same slope for the power-law as well as the high-energy cut-off.
On the other hand, in Fig.~\ref{xspec_badcomp}, we show two cases in which the models are quite different from \texttt{MoCA} results, specifically for a slab corona with $\tau = 5$ and $kT_e = 50$ keV in the upper panel and for a spherical corona with $\tau = 4$ and $kT_e = 20$ keV in the lower panel. In the latter case we performed the comparison with \texttt{Nthcomp} as well.
In the top panel of Fig.~\ref{xspec_badcomp} \texttt{compTT} reproduces the slope observed in MoCA while \texttt{compPS} do not.  While \texttt{compTT} supports such high values for the optical thickness, in \texttt{compPS} we had to force the insertion of the parameter if optical thickness
was higher than 1.5, therefore it is not surprising that it cannot properly reproduce the spectral shape in this regime. Both models however create a bump around 200 keV which is not observed in MoCA. In the lower panel of the figure \texttt{compPS} reproduces the slope below 10 keV while the other two models do not. However, none of the three models reproduce the high-energy cut-off obtained with MoCA.

\subsection{Klein-Nishina effects on the spectrum}

\begin{figure}
\includegraphics[width=88mm,clip=]{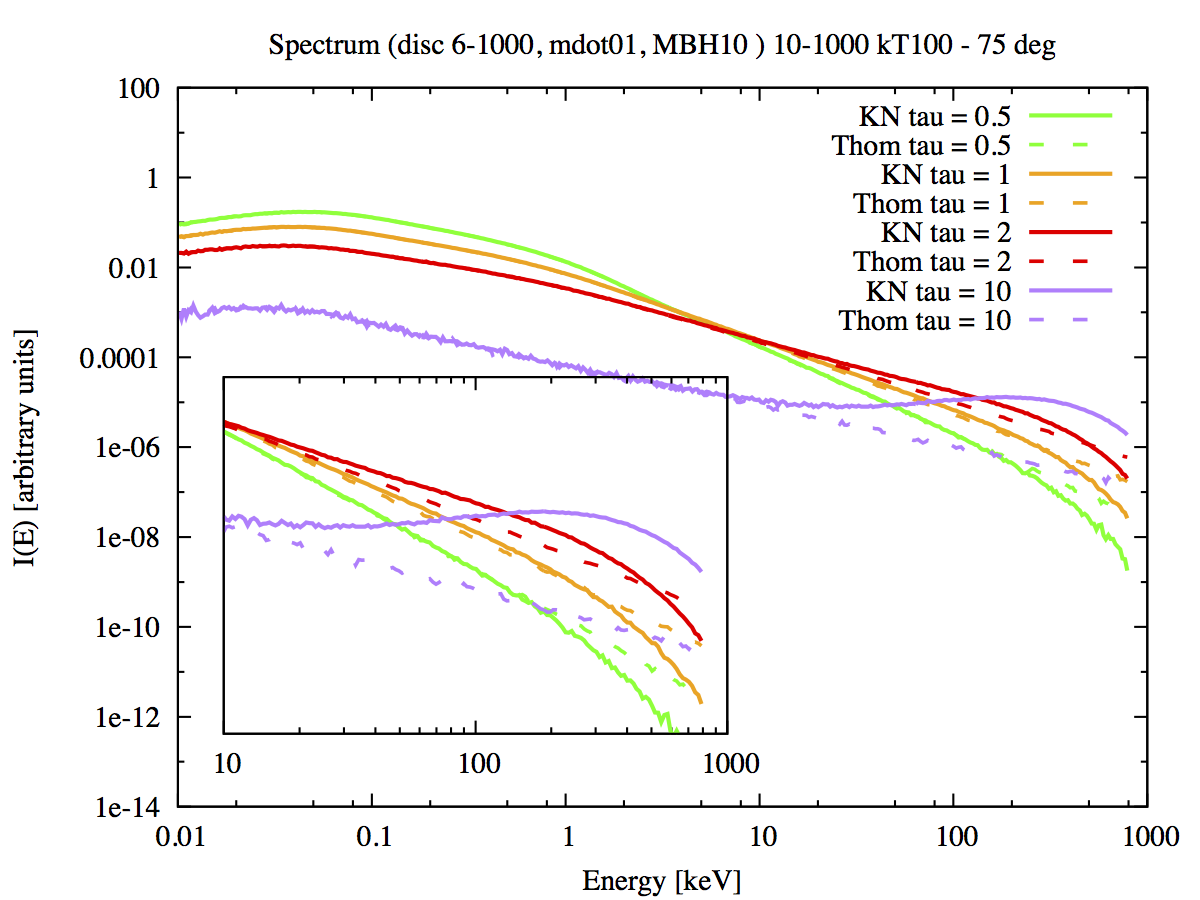}
\caption{\label{spectrum_KNvsThom} Comptonisation spectrum from a slab corona with $kT_e = 100$ keV and increasing $\tau = 0.5,1,2,10$ and for an inclination of 75 degrees. Solid and dashed lines correspond to K-N effects on and off, respectively. In the box we zoomed the high-energy band of the spectra.}
\end{figure}

\begin{figure}
\includegraphics[width=88.mm,clip=]{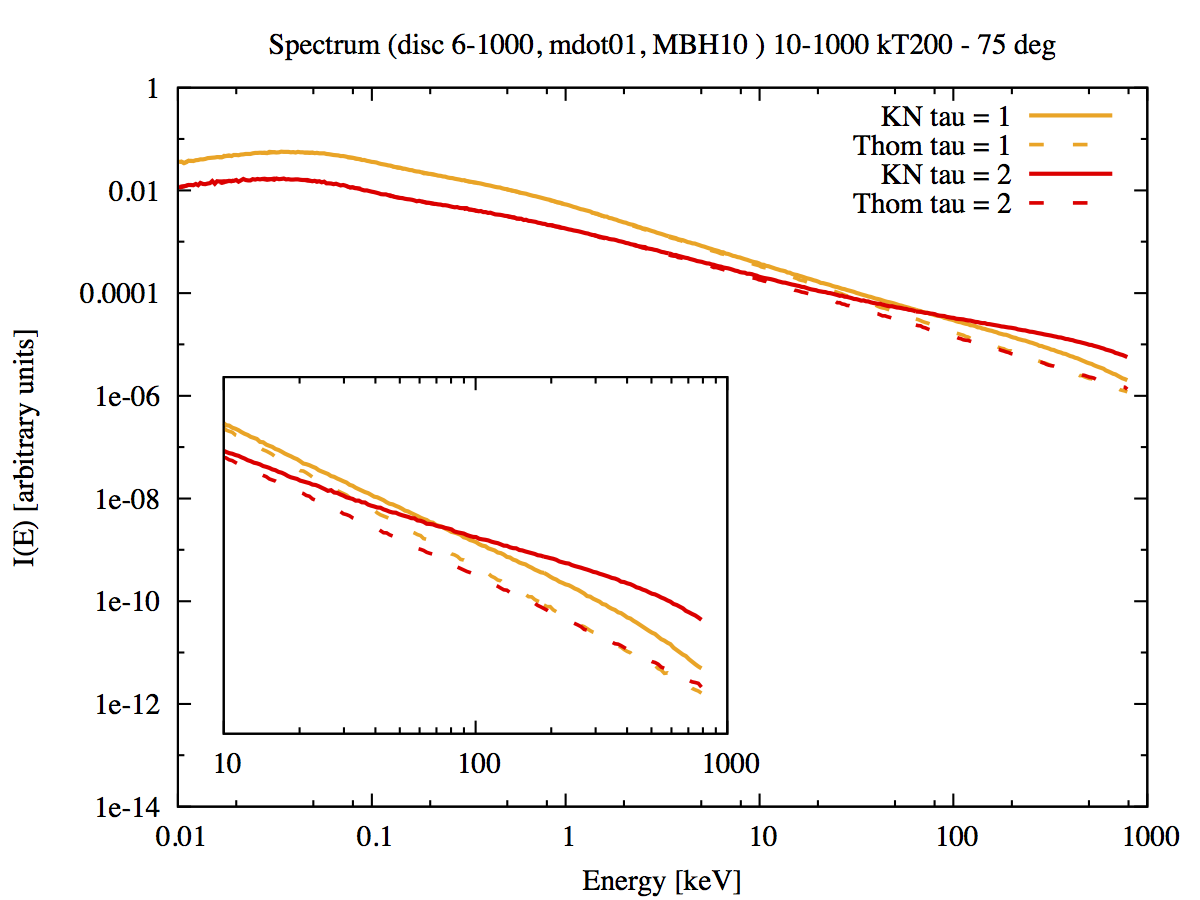}
\caption{\label{spectrum_KNvsThom_kT200} Comptonisation spectrum from a slab corona with $kT_e = 200$ keV and $\tau = 1,2 $ and for an inclination of 75 degrees. Solid and dashed lines correspond to K-N effects on and off, respectively. In the box we magnified the high-energy band of the spectra.}
\end{figure}

In Section 2.3 we mentioned that \texttt{MoCA} uses Klein-Nishina cross-section and scattering angle distribution which may affect the shape of Comptonisation spectrum with respect to the Thomson case, especially at high inclinations where we can look at the photons travelling trough the optically thick slab corona. 
When the energy of the photon in the reference frame of the electron is negligible with respect to the rest mass energy of the electron, all our formulae naturally tends to Thomson.
However, in order to test the discrepancy between our approach and the use of Thomson approximation, we forced our code to use Thomson scattering angle distribution, $\Theta_{sc}$ (Eq. ~\ref{thetasc_Thom} instead of Eq.~\ref{thetasc_KN}), Thomson cross-section (which affects the MFP, Eq. ~\ref{finspace}) and we considered elastic scattering in the reference frame of the electron (instead of Compton scattering, Eq.\ref{compton}), regardless of the energy of the photons. 
In Fig.~\ref{spectrum_KNvsThom} we show the spectra obtained by Comptonisation from a slab with $kT_e = 100$ keV and increasing $\tau = 0.5,1,2,$ and 
Solid and dashed lines correspond to Klein-Nishina and Thomson approximation, respectively. In the box inside the figure we zoomed the high energy band.
We notice that the discrepancies start to appear around 10 keV with the spectra produced in Thomson approximation showing a pure power law profile without any curvature leading to an high-energy cut-off.
When $\tau \le 1$ (green and orange lines) Thomson spectra are harder than K-N spectra. For $\tau = 2$ (red lines) we see the opposite effect with Thomson spectra being softer than K-N spectra and for the extreme $\tau = 10$ case (purple lines) the Comptonisation bump around the thermal energy of the corona as well as the cut-off are completely missing in Thomson spectra.
We tested also the more energetic case of a slab corona with $kT_e = 200$ keV for $\tau = 1, 2$ as shown in Fig.~\ref{spectrum_KNvsThom_kT200}.
In this case we see a larger discrepancy which is much more noticeable even for $\tau = 1$ (orange lines). For $\tau = 2$ the differences are relevant even below 100 keV which can be important to consider when using, for example, NuSTAR data up to 80 keV in order to measure the high-energy cut-off and therefore the thermal energy of the corona.
\cite{Beheshtipour2017} performed a similar test but for a very compact and thick spherical corona lying within 15 $r_g$ from the central object and their spectra are in the form of histograms. However,  they found a qualitatively similar result with Thomson spectra being steeper than K-N ones (model 1 VS model 3 in Fig.4 of their paper).
We defer the analysis of K-N effects on the polarisation to Section 4.4. 

\subsection{Effect of coronal size on the spectrum}
\begin{figure}
\includegraphics[width=88.mm,clip=]{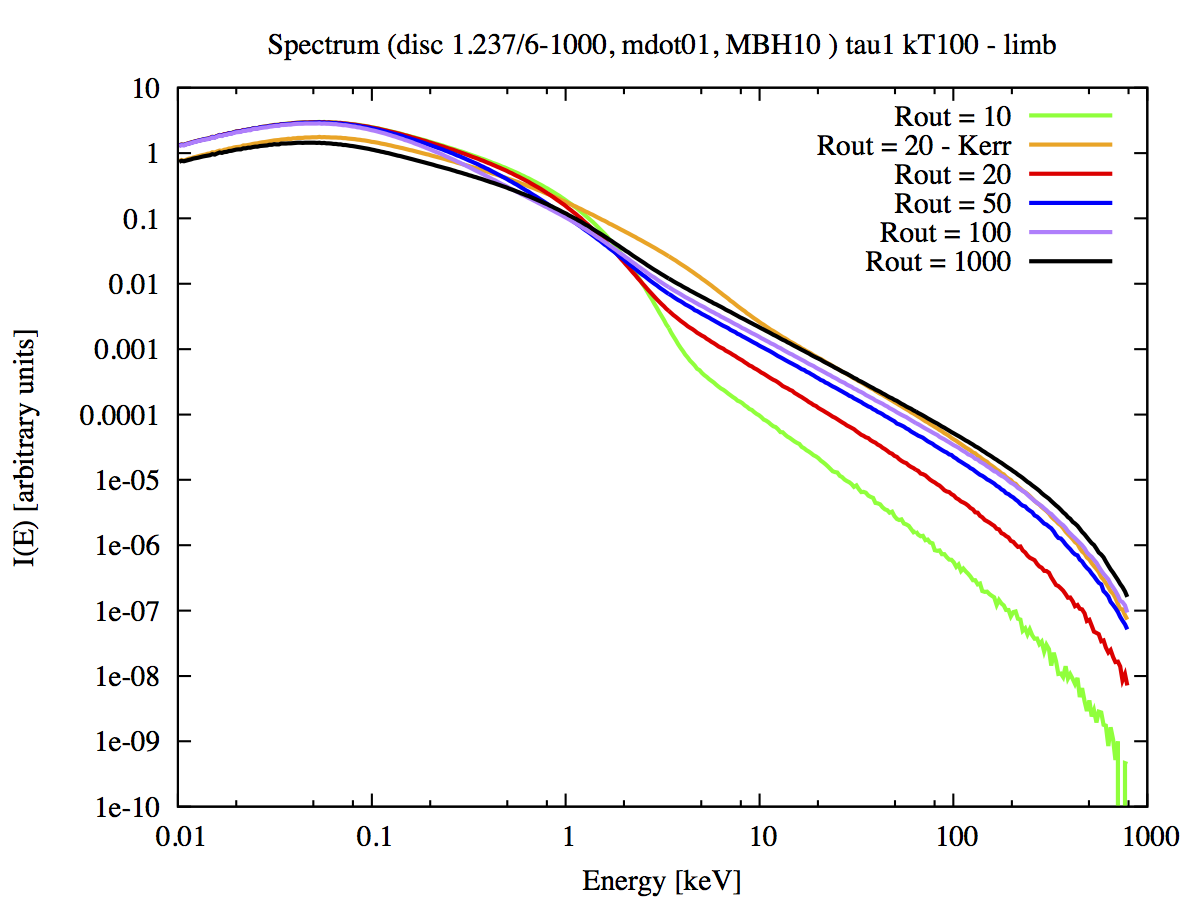}
\caption{\label{spectrum_size} Comptonisation spectrum from a corona with $\tau = 1$ and $kT_e = 100$ keV. The black line correspond to the extended slab corona covering the whole disc up to $1000 \, r_g$. Coloured lines show the spectra as we shrink the outer radius of the corona until we reach a compact spherical corona with radius $10 \, r_g$ (green line). Orange line represents a special case in which we reduced the ISCO to the value corresponding to a Kerr BH (see text for more details).}
\end{figure}

In the previous sections we focussed our attention to the case of extended coronae covering the whole accretion disc. This allowed us to perform a comparison between MoCA and the Comptonisation models available in \texttt{XSPEC}. 
However, as said before, the corona is expected to be very compact within few gravitational radii from the central compact object and at such small distance, GR
effects cannot be neglected and the difference between observed and intrinsic energy can be as large as two to eight times depending on the geometry and size of the corona as well as its inclination \citep{Tamborra2018}.
Nonetheless, even without GR effects, the size of the corona has an impact on both the spectrum and the polarisation signal just from a geometrical point of view.
In Fig.~\ref{spectrum_size} we show the spectra obtained for different sizes of the corona. Black line is the extended slab corona we described in detail in the previous section with height 10 $r_g$ covering the whole disc from the ISCO, 6 $r_g$,  up to 1000 $r_g$. We kept the same height and ISCO and we shrunk the outer radius to 100, 50, 20 and 10 $r_g$ as indicated inside the figure; in all cases the optical depth (defined by the fixed height) is $\tau = 1$ and the thermal energy is $kT_e = 100$ keV.
It can be seen that the spectrum softens with decreasing size of the corona.
Photons travelling tangentially to the disc will see a thinner corona with respect to the case of an extended corona (black line) and multi-scattered radiation, which dominates at higher energies, has less probability of being produced. 
For the corona with outer radius 20 $r_g$ (red line) we considered also the Kerr case with the ISCO at 1.237 $r_g$ (orange line). In the latter case the emission originates closer to the central object and the effective optical depth seen by the photons is larger than the Schwarzschild case. The resulting spectrum becomes harder and comparable with more extended coronae.
We defer the effects of coronal size on the polarization signal to Section 4.5.

\section{Polarisation signal}
One of the key features of \texttt{MoCA} is that it includes polarisation. Comptonisation (like any scattering process) induces a polarisation signature in scattered radiation
whose net result depends, mainly, on the (inclination at which we look at the) geometry of the scattering material and less prominently on the relative energy of the seed photons with respect to the energy of the scattering material. Radiation scattered several times in a medium will show a polarisation perpendicular to the plane in which it scattered.
In Section 4.1 we will talk in general about how the scattering-induced polarisation is build in order to understand our results which will be shown in Section 4.2. In Section 4.3 we compare the polarisation signal for the two geometries we are taking into account in this paper and in Section 4.4 we consider the Klein-Nishina effects on the polarisation signal similarly to what we have shown on the spectrum in Section 3.1. Finally, in Section 4.5 we show how the coronal size influences the polarisation signal.

\subsection{Understanding polarisation}
\begin{figure}
\includegraphics[width=88.mm,clip=]{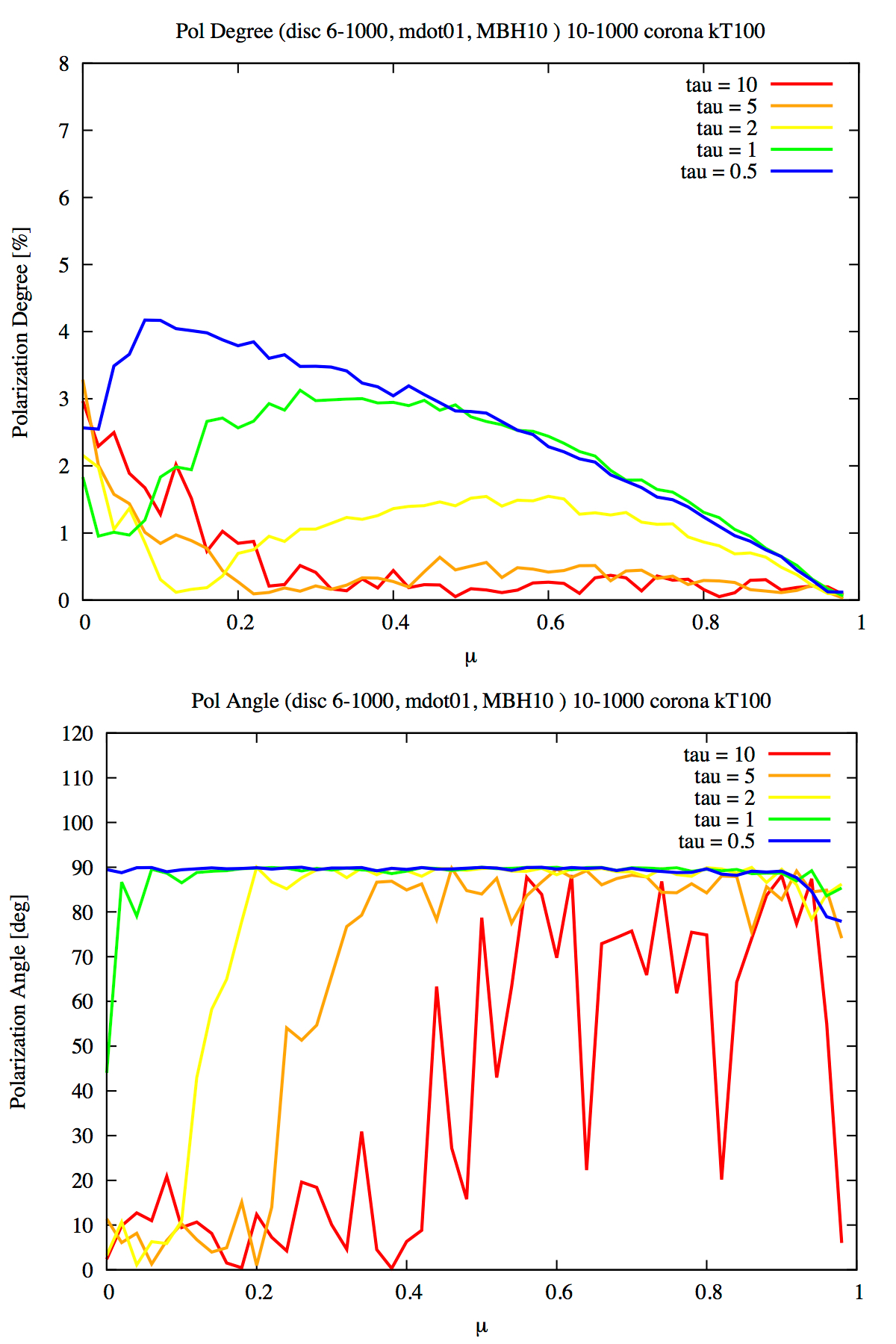}
\caption{\label{POL_mu_disc_multitau} Polarisation signal as a function of the cosine of inclination with respect to the spin-axis for different values of $\tau$ for a slab corona with $kT_e = 100$ keV. \textit{Upper panel:} Polarisation degree. \textit{Lower panel:} Absolute value of the corresponding polarisation angle.}
\end{figure}

\begin{figure}
\includegraphics[width=88.mm,clip=]{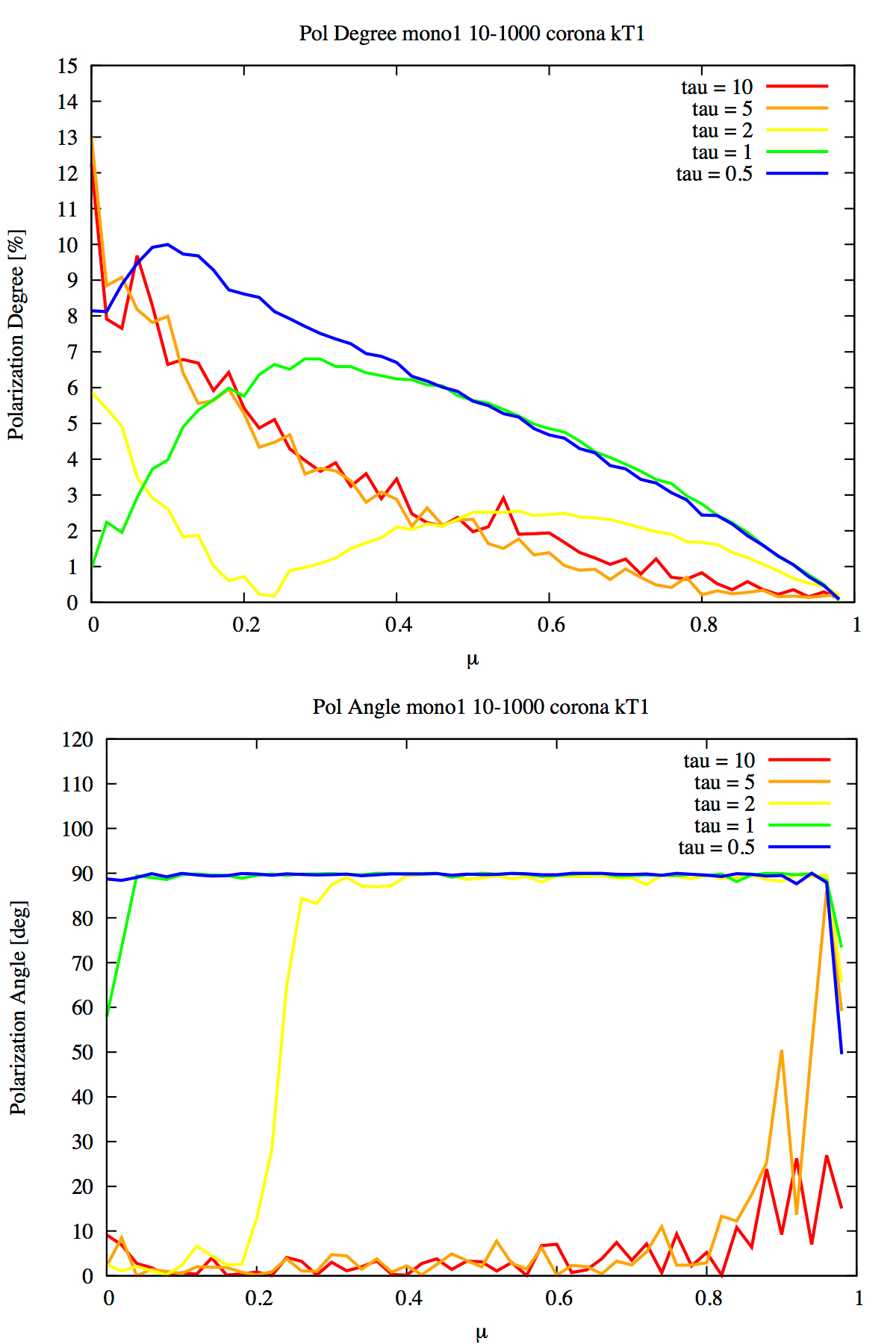}
\caption{\label{POL_mu_test_multitau} Polarisation signal as a function of $\mu$ and for different values of $\tau$, similar to Fig.~\ref{POL_mu_disc_multitau}  but for the test case of a point-like source emitting at 1 keV at the centre of a  a slab corona with $kT_e = 1$ keV. \textit{Upper panel:} Polarisation degree. \textit{Lower panel:} Absolute value of the corresponding polarisation angle.}
\end{figure}

Before showing our results, we note that for the axisymmetric geometries we chose (see Fig.~\ref{geometries}) we expect the polarisation angle to be 0 or 90 degrees with only the Stokes parameter Q, which describes these two polarisation state, playing any role while the Stokes parameter U, describing polarisation angles at 45 degrees with respect of the symmetry axis, being irrelevant. This might not have been the case if GR effects were 
included due to the rotation of the polarisation vector along geodesics in a strong gravitational field which might lead to a net polarisation signal with intermediate  angles even for axisymmetric geometries.
Also, the polarisation vector is a pseudo-vector with defined orientation but not defined sign so when Eq.~\ref{piechi} is applied to derive the polarisation angle, $\chi$, +90 and -90 degrees are equivalent and we then take the absolute value.
Finally, we address the way in which we build the Stokes parameters in MoCA.
A polarisation angle of 0 degrees corresponds with having the photons polarised perpendicularly to the spin axis (or parallel to the plane of the disc) and from here on we refer to this polarisation as 'horizontal'. On the other hand, a polarisation angle of 90 degrees correspond to a polarisation parallel to the spin axis and we refer to it as 'vertical'.
In Fig.~\ref{POL_mu_disc_multitau} we show the polarisation signal as a function of the inclination ($\mu = \cos(\theta)$) for different optical thickness of a slab corona with  $kT_e = 100$ keV. The initial radiation coming from the disc is unpolarised (Q,U = 0) and isotropically emitted at any given radius. 
The first thing to notice is that the polarisation signal disappear if we look at the system face-on (i.e. $\mu = 1$) and that is always true, being all the geometries we considered axis-symmetric.
As we move from a completely face-on view to intermediate inclinations, the polarisation signal for a given value of $\mu$ strongly depends on the optical thickness of the scattering material. When the slab corona is thin ($\tau = 0.5$, blue line) or moderately thick ($\tau = 1$, green line) the polarisation is vertical (i.e.  $\chi$ = 90 degrees) because the photons will travel and scatter through the horizontal slab before escaping to the observer.
As the corona becomes thicker we see a transition from vertical to horizontal polarisation if we look at the system edge-on (i.e. $\mu \le 0.4$). 
This is due to the fact that at such high optical depths only the photons travelling upwards (i.e. from the bottom to the top of the slab corona) are able to escape while those travelling along the slab will never reach the observer. The photons which escape the thick coronae and reach the observer looking at the system edge-on will then be horizontally polarised. For $\tau = 10$ (red line) one might expect a Chandrasekhar horizontal polarisation reaching $12\%$ for $\mu = 0$ while in Fig.~\ref{POL_mu_disc_multitau} it reaches only $\sim 3 \%$. 
This is mainly due to the Klein-Nishina regime of the scattering between soft photons emitted from the disc around 0.1 keV for a stellar BH and the highly energetic corona at 100 keV and it is also partially due to the finiteness of the corona (and the fact that the source, the disc, is extended below the scattering material).
We performed some tests with a point source emitting 1 keV isotropic unpolarised radiation at the centre of the slab with the same thermal energy, $kT_e = 1$ keV.
In Fig.~\ref{POL_mu_test_multitau} we show the resulting polarisation signal for this test case, similarly to Fig.~\ref{POL_mu_disc_multitau}.
Qualitatively the results are very similar to the more realistic case of an emitting disc showed previously.
However in Thomson regime we see that for $\tau = 5$ (orange line) and for $\tau = 10$ as well (red line) we already find Chandrasekhar results: the horizontal polarisation increases from $0 \%$ to $\sim 12 \%$ when moving from a face-on to edge-on view. These results are in good agreement with the first Monte Carlo simulations performed on polarisation by \cite{Angel1969} and later by \cite{Dovciak2008}.

\subsection{Polarisation vs energy}

\begin{figure}
\includegraphics[width=88.mm,clip=]{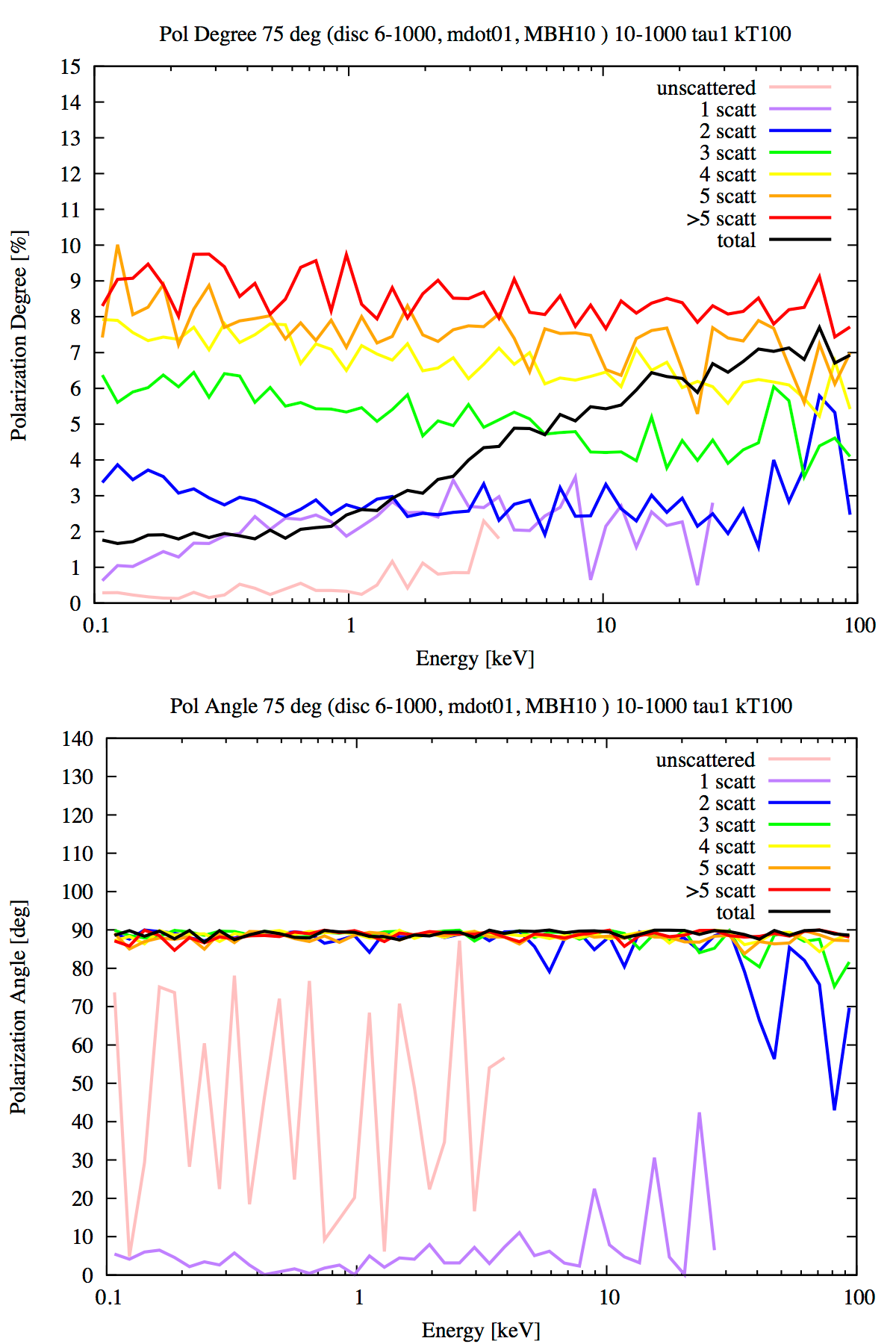}
\caption{\label{POL_en_disc_multi_75} Polarisation signal as a function of the energy for the case of a slab corona with $\tau = 1$ and $kT_e = 100$ keV decomposed in scattering orders as indicated inside the figure. The seed photons produced by the disc are unpolarised and the system is observed at 75 degrees inclination. \textit{Upper panel:} Polarisation degree. \textit{Lower panel:} Absolute value of the corresponding polarisation angle.}
\end{figure}

In Fig.~\ref{POL_en_disc_multi_75} we show the polarisation signal as a function of energy for the case of a slab corona with $\tau = 1$ and $kT_e = 100$ keV decomposed in scattering orders as indicated inside the figure, for an observer at 75 degrees (almost edge-on).
In this first example the seed photons produced by the Shakura-Sunyaev disc are initially unpolarised, as can be clearly seen by the pink line representing the unscattered component reaching the observer at infinity: the polarisation degree is zero, with the polarisation angle being random.

In this scenario we expect to end up with a vertical polarisation ($\chi = 90$ degrees) which reflects the geometry of the scattering material (a moderately thick horizontal slab, where the scatterings happen). The multi-scattered component (red line) formed by photons which traveled the most trough the slab corona becomes dominant at higher energies and as a consequence the total polarisation degree increases (black line).
If we look at the system almost face-on, it is much more symmetric and the expected polarisation degree is very low. We do not show the decomposed polarisation signal in this case but we consider different inclinations later. 
Anyway, if we look at it almost edge-on (Fig.~\ref{POL_en_disc_multi_75}), we can reach 8 \% of polarisation degree at high energies .
It is also interesting to notice the special behaviour of the one-scattered component represented by the purple line in the figure: these photons are collected by an edge-on observer and they experienced only one scattering. The only way it can happen is if the photons are emitted upwards and then scatter at $\sim 90$ degrees which explain the horizontal polarisation (being the scattering plane vertical, or more precisely perpendicular to the disc plane).

\begin{figure}
\includegraphics[width=88.mm,clip=]{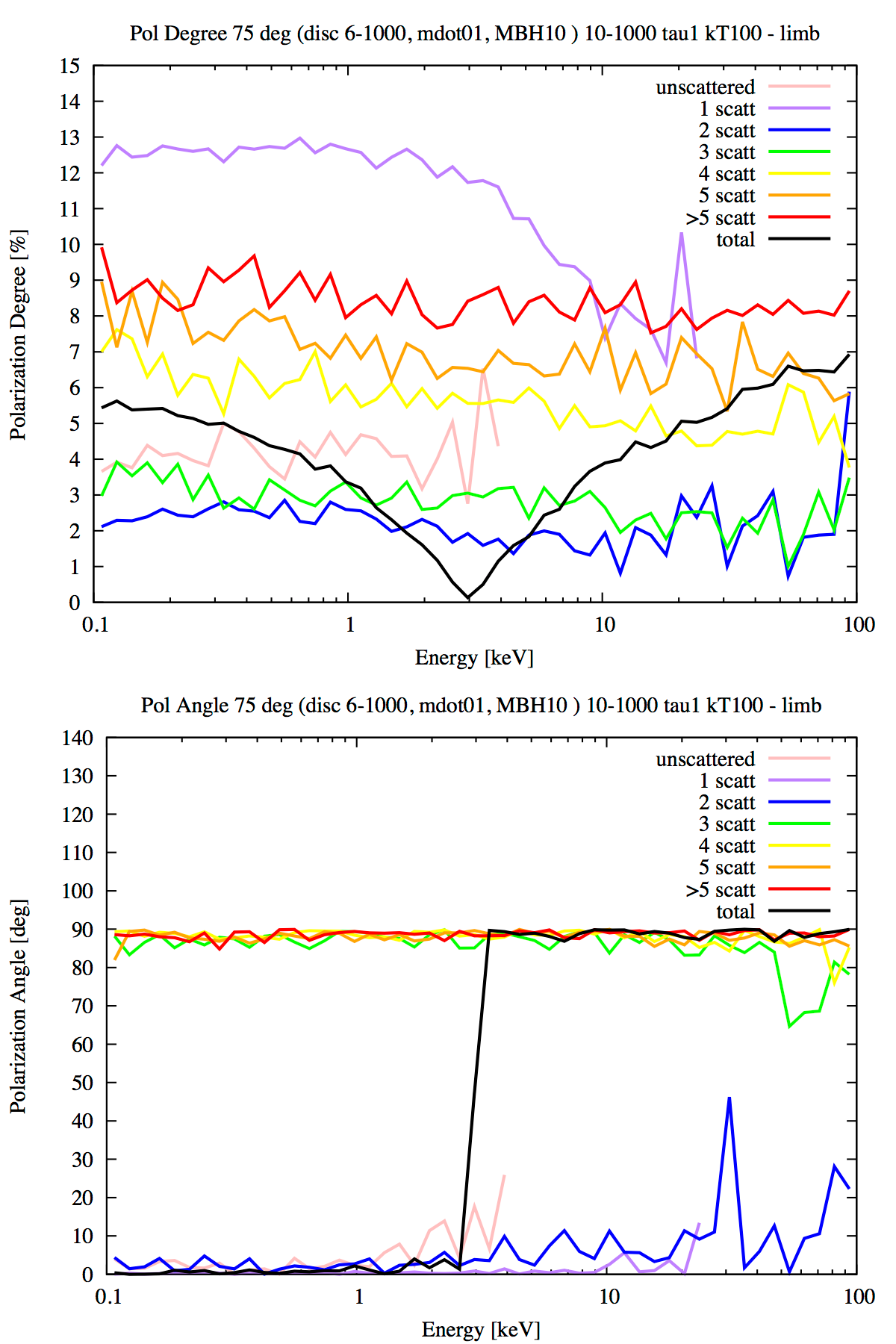}
\caption{\label{POL_en_disc_multi_75_limb} As Fig.~\ref{POL_en_disc_multi_75} but we included limb darkening and the correspondent initial Chandrasekhar polarisation to the seed photons. \textit{Upper panel:} Polarisation degree. \textit{Lower panel:} Absolute value of the corresponding polarisation angle.}
\end{figure}

In the previous section, as well as in the second section, we mentioned the possibility of the disc emitting radiation which might be initially polarised according to Chandrasekhar calculations in the approximation of an infinitely thick disc atmosphere (see Section 2.1). 
In Fig.~\ref{POL_en_disc_multi_75_limb} we show the polarisation signals decomposed in scattering orders, as in Fig.~\ref{POL_en_disc_multi_75} for an observer at 75 degrees inclination. In this second example we gave initial horizontal polarisation up to 12\% to the few (because we included limb darkening as well, as described in Section 2.1) seed photons arising tangentially from the disc, according to Chandrasekhar calculations. 
The unscattered component described by the pink line in the figure shows this initial horizontal polarisation ($\chi = 0$ degrees).
The most important thing to notice is that the initial polarisation completely change the outcome of the total polarisation (black line): at lower energy, where unscattered and few-scattered components dominate we have an horizontal polarisation which then flips to vertical where the photons experiencing three or more scattering 
start to dominate the energy band. For an edge-on view, however, we can observe a quite high polarisation degree of 6-8 \% flipping from horizontal to vertical around 3 keV for this particular case.

\begin{figure}
\includegraphics[width=88.mm,clip=]{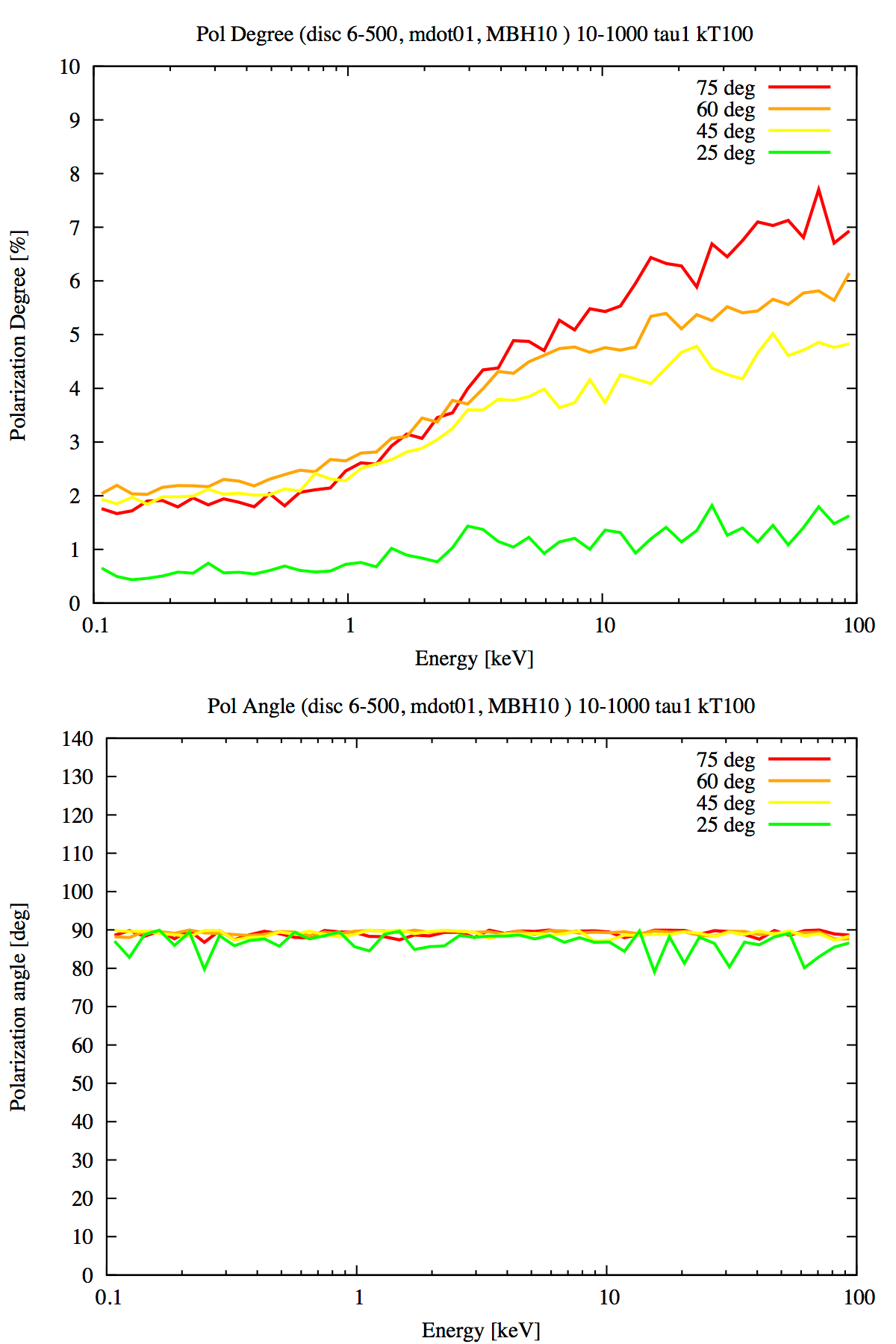}
\caption{\label{POL_en_disc_AOV} Observed polarisation signal for a slab corona with $\tau = 1$ and $kT_e = 100$ keV observed at different inclinations as indicated inside the figure. The seed photons are unpolarised and isotropically emitted at a given radius. \textit{Upper panel:} Polarisation degree. \textit{Lower panel:} Corresponding polarisation angle.}
\end{figure}

\begin{figure}
\includegraphics[width=88.mm,clip=]{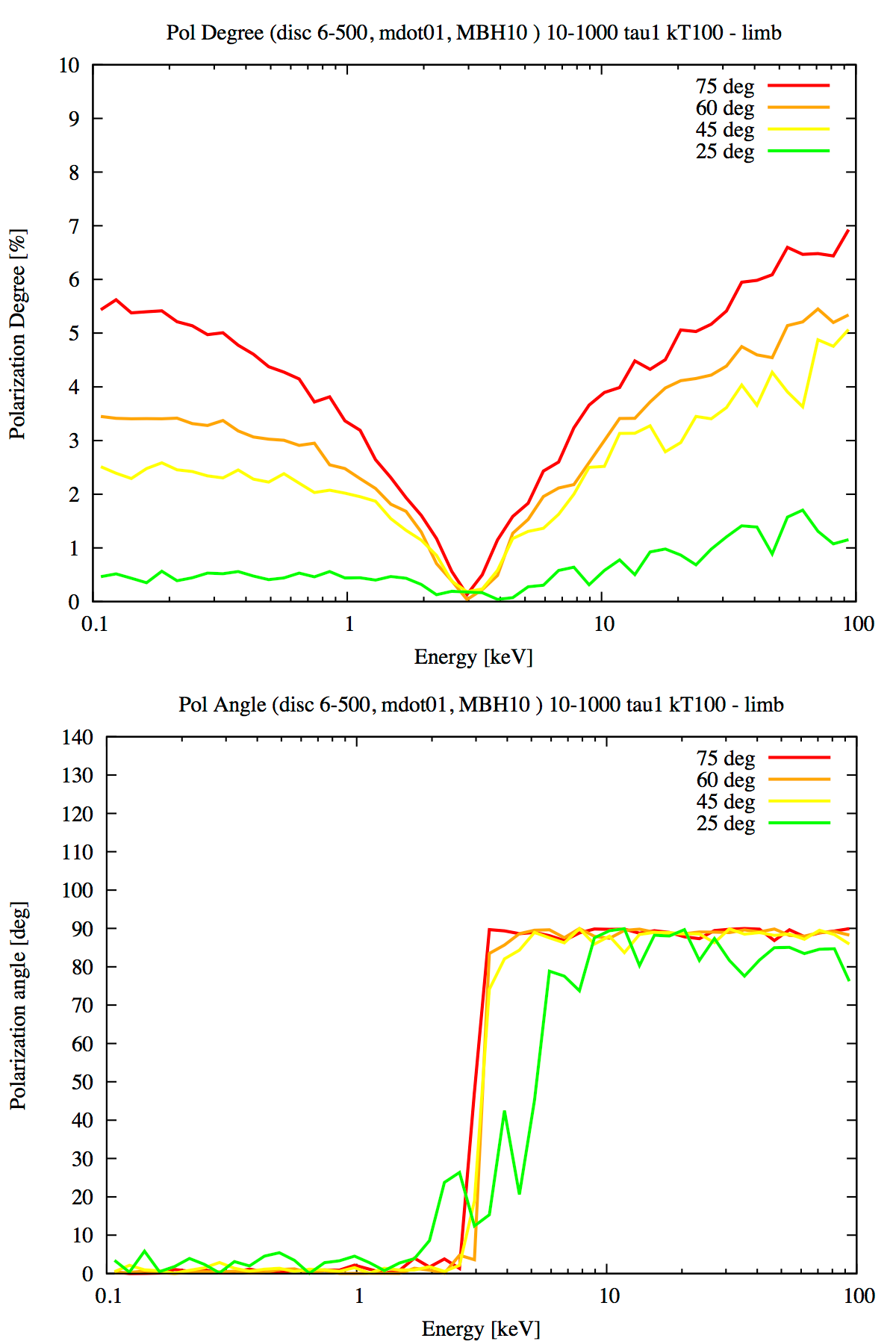}
\caption{\label{POL_en_disc_AOV_limb} As Fig.~\ref{POL_en_disc_AOV} but we applied limb darkening and the corresponding initial polarisation to seed photons according to Chandrasekhar calculations. \textit{Upper panel:} Polarisation degree.  \textit{Lower panel:} Absolute value of the corresponding polarisation angle.}
\end{figure}

Once again, thanks to the single-photons approach we use in MoCA we can access information such as the scattering order of the photons, which allows us to deeply understand the complex mechanism in play to form the polarisation signal.
In Fig.~\ref{POL_en_disc_AOV} we show the observed polarisation signal for our template case of a slab corona with $\tau = 1$ and $kT_e = 100$ keV observed at different inclinations as indicated inside the figure. The seed photons are unpolarised and isotropically emitted at a given radius.
In Fig.~\ref{POL_en_disc_AOV_limb} we show the same signal as in Fig.~\ref{POL_en_disc_AOV} but this time we include limb darkening and the seed photons have initial Chandrasekhar polarisation.

The two signals are completely different. If the seed photons are unpolarised we expect to observe only vertical polarisation (lower panel of Fig.~\ref{POL_en_disc_AOV}) with the polarisation degree being higher as we move towards higher inclinations and reaching 6-7 \% for 75 degrees.
On the other hand, if the seed photons are initially polarised as for Chandrasekhar optically thick atmospheres, we see a characteristic flip in the polarisation angle from horizontal to vertical when the multi-scattered (three or more scatterings) component begins to dominate the energy band which in this particular case happens around 3 keV. 
The results shown in Fig.~\ref{POL_en_disc_AOV_limb} are very similar to the findings of \cite{Schnittman2010} (Fig. 3 in their paper) even though they use a different approach and their code includes all GR effects as well (but in Thomson regime). This shows that GR effects are not playing a dramatic role in this scenario. However it is not surprising considering that in both works the Comptonisation (and the relative polarisation signal) comes from very extended coronae and GR effects are expected to be negligible at several $r_g$ distance from the central compact object. 

\subsection{Polarisation for different geometries}
\begin{figure}
\includegraphics[width=88.mm,clip=]{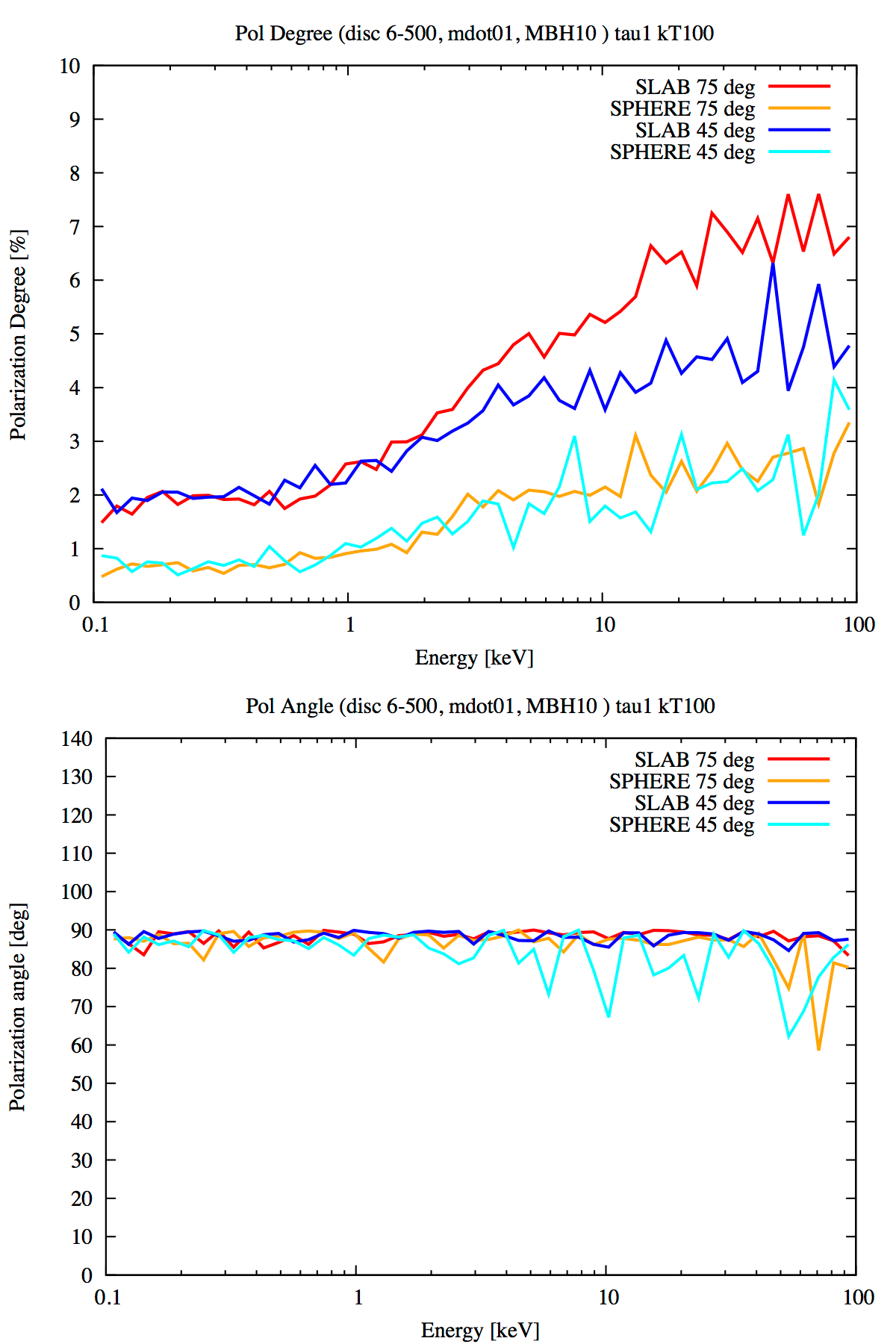}
\caption{\label{POL_geom_AOV} Comparison between the polarisation signal for a slab and a spherical corona (both with $\tau = 1$ and $kT_e = 100$ keV) for two different inclinations as indicated inside the figure. The seed photons are isotropically emitted from a given radius and they are unpolarised. \textit{Upper panel:} Polarisation degree. \textit{Lower panel.} Absolute value of the corresponding polarisation angle.}
\end{figure}

\begin{figure}
\includegraphics[width=88.mm,clip=]{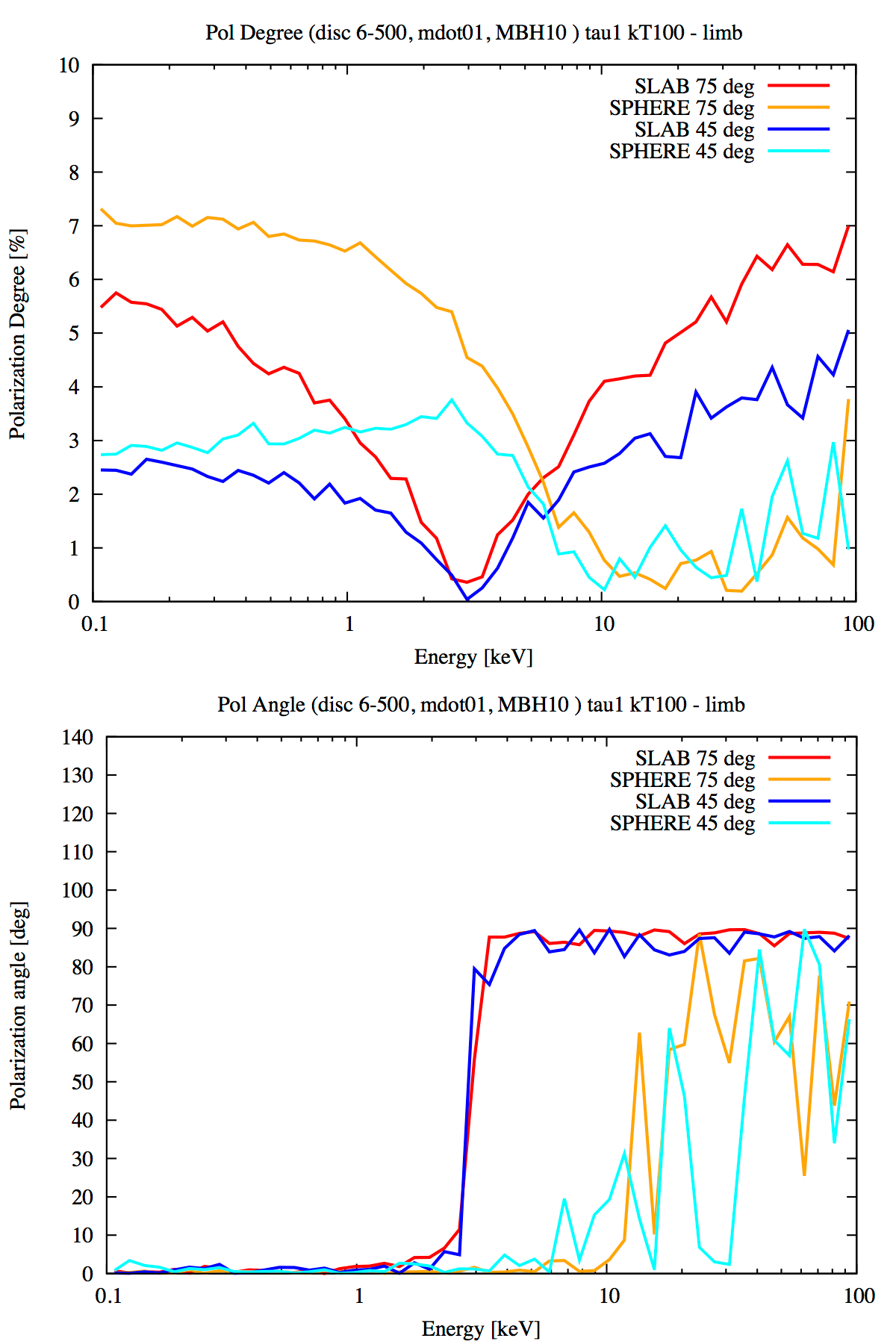}
\caption{\label{POL_geom_AOV_limb} As Fig.~\ref{POL_geom_AOV} but with limb darkening and the corresponding initial polarisation for the seed photons according to Chandrasekhar calculations. \textit{Upper panel:} Polarisation degree. \textit{Lower panel.} Absolute value of the corresponding polarisation angle.}
\end{figure}

Up to now, we have focussed entirely on the case of a slab corona because it represents a much more interesting and instructive case when studying the polarisation signal, appearing completely symmetric if observed face-on and very axisymmetric if observed edge-on.
In the trivial tests we performed with a point-like source inside a sphere, being the system completely symmetric  by definition from any angle of view, the polarisation is always zero regardless the optical thickness or the energy of the spherical Comptonising medium.
In the more realistic case of a hemispherical corona covering the whole disc (Fig.~\ref{geometries}), the presence of the disc (and hence the base of the hemisphere) introduces the only asymmetry which in the end makes the hemisphere behave like a slab corona but much less efficient in term of Comptonisation (because the optical thickness in the slab is defined vertically while in the hemisphere is defined radially) and polarisation (because, of course, the slab deviates more from spherical symmetry).
In Fig.~\ref{POL_geom_AOV} we show the comparison between the observed polarisation signal for a sphere and a slab (both with $\tau = 1$ and $kT_e = 100$ keV) for two different inclinations as indicated inside the figure. The seed photons are unpolarised.
In all cases the observed polarisation is vertical (lower panel of Fig.~\ref{POL_geom_AOV}) but for the slab we have a larger degree of polarisation, which can reach 4-5 \% at higher energies (blue line) versus a 2-3 \% for the sphere (cyan line) if we look at the system at 45 degrees inclination.
For a higher inclination of 75 degrees, however, we can have up to 7 \% polarisation degree for the slab (red line) while for the much more symmetrical sphere it remains around 2-3 \% (orange line).
In Fig.~\ref{POL_geom_AOV_limb} we show a similar plot but this time we included limb darkening and the seed photons has initial Chandrasekhar polarisation.
In this case the slab corona shows the typical transition between horizontal and vertical polarisation explained in the previous section (red and blue lines for the two inclinations). For the sphere (orange and cyan lines), however, this transition is almost not happening: at lower energies up to $\sim 10$ keV unscattered and few-scattered photons completely dominate and we see horizontal polarisation which can reach 7-8 \% if seen at 75 degrees inclination and 3-4 \% at 45 degrees. Above 10 keV we have multi-scattered components even in the sphere but the vertical polarisation struggles to reach more than 1 \% polarisation degree for both inclinations.

\begin{figure}
\includegraphics[width=88.mm,clip=]{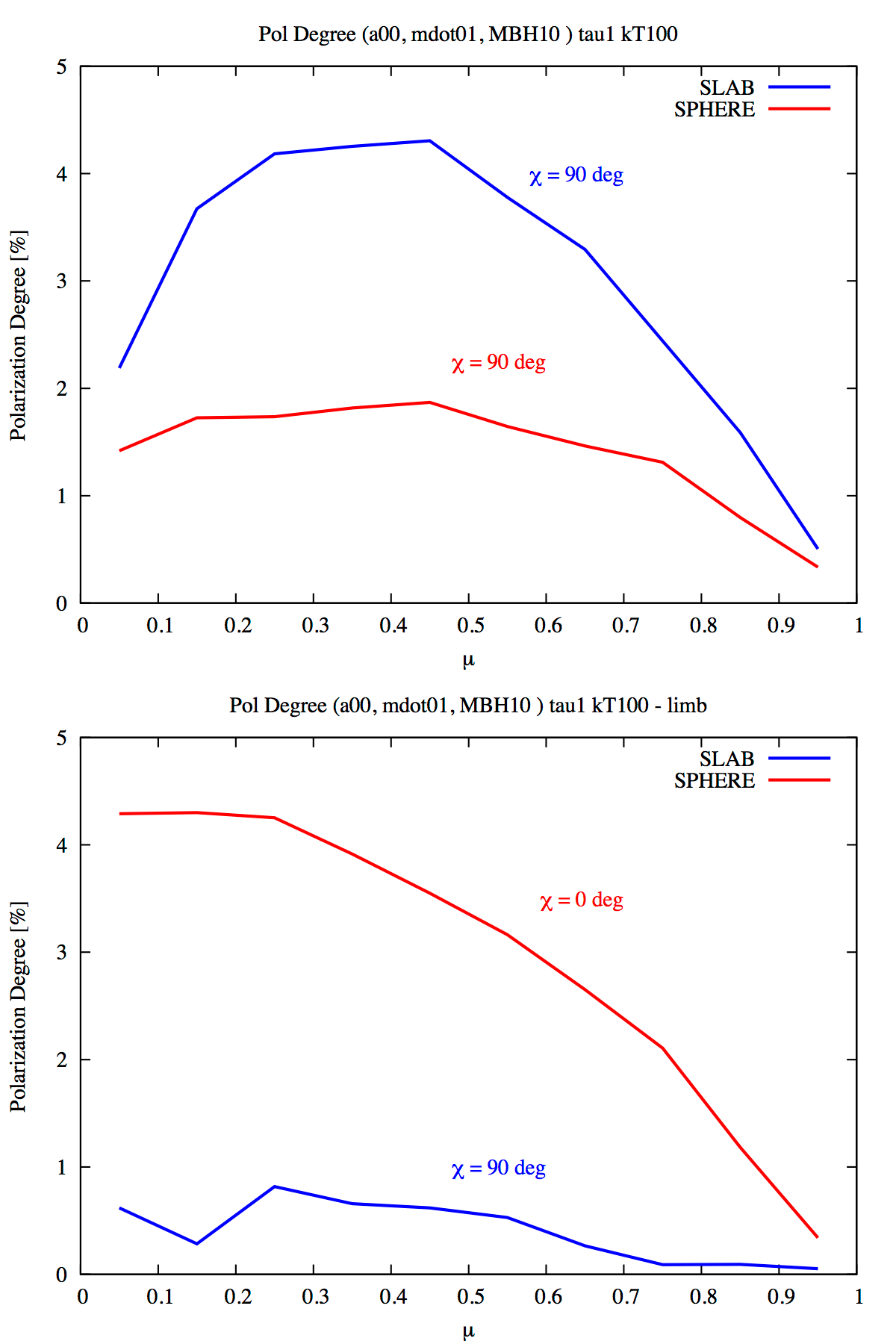}
\caption{\label{POL_geom_IXPE} Polarisation degree integrated between 2 and 8 keV (IXPE energy band) for the two coronal geometries as indicated inside the figure as a function of the inclination (10 $\mu$ bins). \textit{Upper panel:} Case of initially unpolarised photons. \textit{Lower panel:} Chandrasekhar scenario. We report the absolute values of the corresponding polarisation angle, $\chi$, inside the figures.}
\end{figure}

The spectral resolution of the next generation X-ray polarimeters such as the one proposed for the IXPE mission, estimated around 1 keV at 6 keV, will be essential to discriminate between coronal geometries especially in the limb darkening scenario (Fig.~\ref{POL_geom_AOV_limb}). 
However, even integrating the polarisation signal in the IXPE polarimeter band between 2 and 8 keV, the two geometries we consider in this work can still be distinguishable.
In Fig.~\ref{POL_geom_IXPE} we show the 2-8 keV signal in ten bins of inclination for the two geometries as indicated inside the figure and for the case of unpolarised seed photons in the upper panel and for the Chandrasekhar case in the lower one. For the initially unpolarised scenario, for both geometries the polarisation is vertical (as indicated by the value of the polarisation angle, $\chi$, inside the figure equal to 90 degrees) but the slab shows a much higher degree of polarisation at any inclination.
Even more interesting is the same comparison but for the case of initially Chandrasekhar polarised radiation shown in the lower panel of Fig.~\ref{POL_geom_IXPE}.
In this case for the slab (blue line) we still have vertical polarisation but the polarisation degree is quite low, below 1 \%. The reason is that between 2 and 8 keV we are integrating the region in energy where we have the flip of the polarisation angle (see Fig.~\ref{POL_geom_AOV_limb}) so the average polarisation degree will be very low. On the other hand the polarisation angle, by definition, does not take into account the photon counts (Eq.~\ref{piechi}) so because for most of the energy bins between 2 and 8 keV we have vertical polarisation for any inclinations, the integrated polarisation angle will still be vertical.
For the sphere (red line) instead we have a quite strong horizontal polarisation as expected by integrating between 2 and 8 keV (see orange and cyan lines in Fig.~\ref{POL_geom_AOV_limb}). For both scenarios, at least for the parameters we have chosen in the simulations, IXPE polarimeter will be able to discriminate between the two geometries even integrating the signal in the whole operating energy band.

\subsection{Klein-Nishina effects on the polarisation}
\begin{figure}
\includegraphics[width=88.mm,clip=]{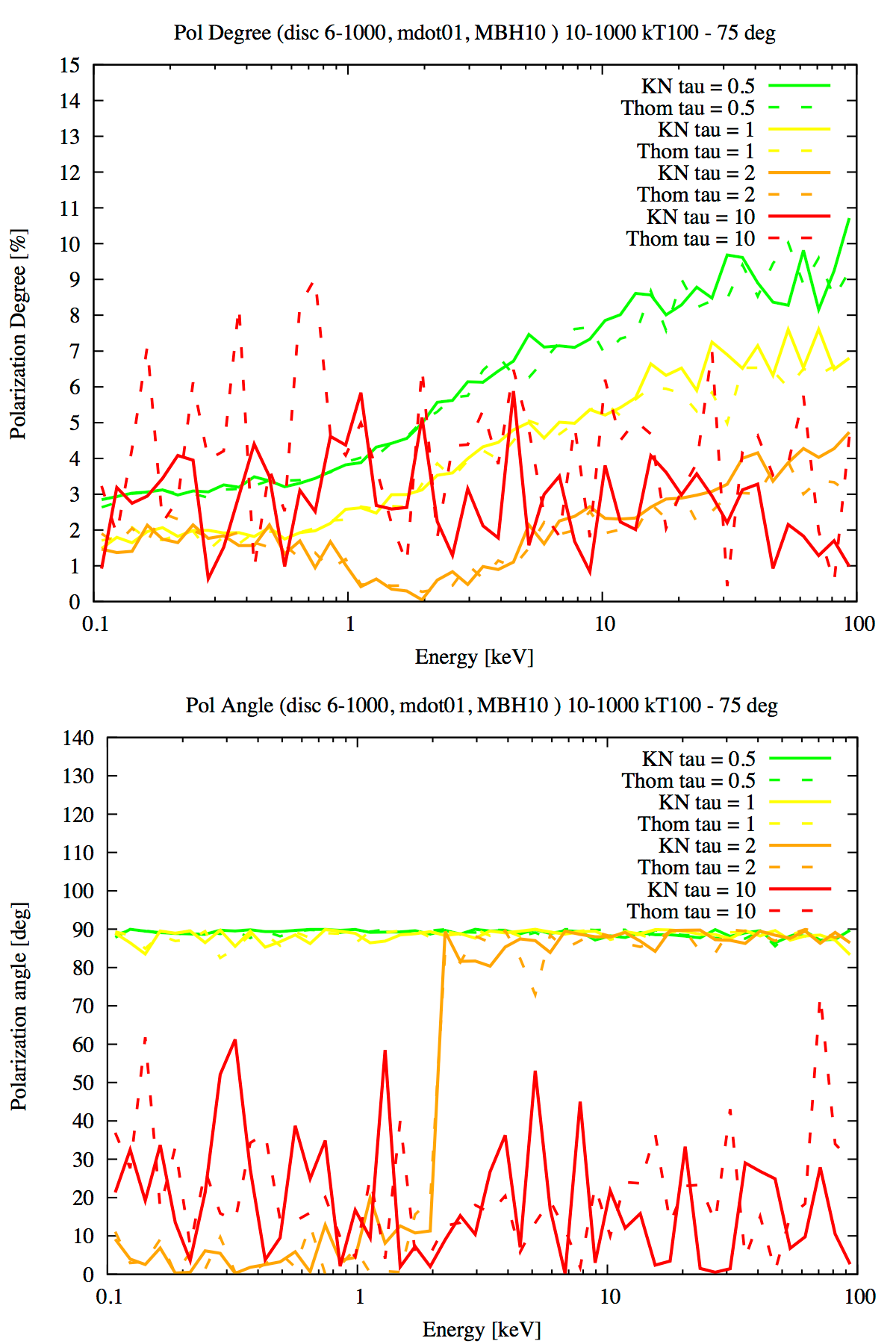}
\caption{\label{POL_KNvsThom} Polarisation signal from a slab corona with $kT_e = 100$ keV and increasing $\tau = 0.5,1,2,10$ and for an inclination of 75 degrees. Solid and dashed lines correspond to K-N effects on and off, respectively. \textit{Upper panel:} Polarisation degree. \textit{Lower panel:} Absolute value of the corresponding polarisation angle.}
\end{figure}

In Section 3.1 we describe how we tested the effect of using the Klein-Nishina cross-section, scattering angle distribution and Compton energy shift on the spectra and we found that the deviations from the Thomson approximation were slightly visible below 100 keV (for not extreme optical depths or very hot coronae). We performed those tests for the polarisation as well and we show them in Fig.~\ref{POL_KNvsThom} for a slab corona with $kT_e = 100$ keV and increasing $\tau = 0.5,1,2,10$ and for an inclination of 75 degrees; solid and dashed lines correspond to K-N effects on and off, respectively, and the seed photons are initially unpolarised.
As expected, the polarisation signal does not change much. The K-N effects are well visible above 100 keV which is above the band in which we plotted the polarisation signal. We do not find it useful to report the polarisation signal above 100 keV because i) the noise due to low photon counts will dominate with respect to the small variation on the polarisation signal in the two regimes and ii) also because the next generation polarimeters will work in the soft X-rays.
Nonetheless one can imagine a scenario in which the seed photons are much less energetic (for example the UV photons arising from the accretion disc of an AGN) and the corona having a thermal energy of few tens of keV. In this case we expect to see a (probably still small) difference in the polarisation signal in the IXPE band for the two regimes but we defer this study to future papers focussed on the exploration of the coronal parameters space in AGN.
\cite{Beheshtipour2017} found an interesting difference in the polarisation signal when comparing the proper K-N treatment VS the Thomson one but they compared models which were giving the same spectral index in the 2-10 keV energy band leading to an optical depth per proper length which is more than two times larger for the Thomson case with respect to the K-N one (Fig.6 in their paper).

\subsection{The effect of coronal size on the polarisation}

\begin{figure}
\includegraphics[width=88.mm,clip=]{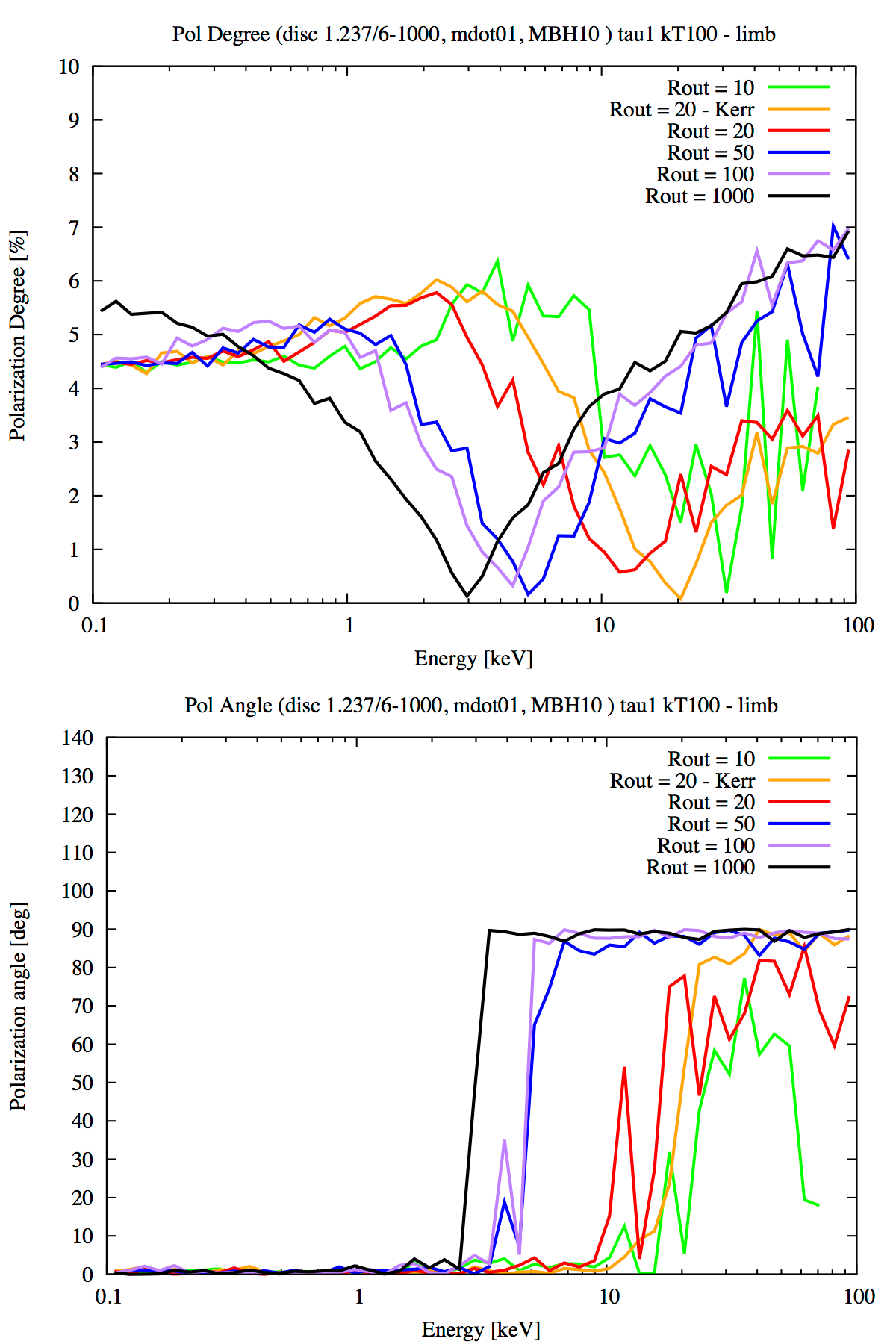}
\caption{\label{POL_size_limb} Polarisation signal from a corona with $\tau = 1$ and $kT_e = 100$ keV observed at 75 degrees inclination with seed photons initially polarised according to Chandrasekhar calculations. The black line correspond to the extended slab corona covering the whole disc up to $1000 \, r_g$. Coloured lines show the signal as we shrink the outer radius of the corona until we reach a compact spherical corona with radius $10 \, r_g$ (green line). Orange line represents a special case in which we reduced the ISCO to the value corresponding to a Kerr BH (see text for further detail).}
\end{figure}

In Section 3.2 we showed how the spectrum changes as we shrink the extended slab corona covering the whole disc to a compact spherical corona with radius $10 \, r_g$ around the central object. The main effect from the spectroscopical point of view was a significant reduction of the Comptonising power of the corona and the result was a softer spectrum. 
On the other hand the polarisation signal is much more sensitive to the geometry of the scattering material.
In Fig.~\ref{POL_size_limb} we show the polarisation signal at 75 degrees inclination for different truncation radius of the slab corona similarly to Fig.~\ref{spectrum_size} for the case of Chandrasekhar initially polarised photons.

We see that as we shrink the corona the flip in the polarisation angle moves to
higher energies. When we reach the spherical corona (green line) the flip happens above 10 keV and we have a very low vertical polarisation of few \%.
Comparing the signal in Fig.~\ref{POL_size_limb} with the polarisation signal we showed for the two geometries in Section 4.3, shrinking the corona is equivalent to moving form the red to the orange line in Fig.~\ref{POL_geom_AOV_limb} with the difference that the spherical corona in this case is very compact instead of covering the whole disc and the resulting signal is much more noisy.
For a truncation radius of $20 \, r_g$ (red line) we included also our special Kerr case in which we set the ISCO at $1.237 \, r_g$ (orange line). While the spectrum in the Kerr case was much harder and comparable to more extended coronae, the polarisation signal is very similar with the flip in the polarisation angle moved at slightly higher energies. This is not surprising because the geometry of the corona is quasi-spherical in both cases and the only difference is that for the Kerr case the innermost and more active region of the disc is well inside the corona making the corona behave as if it was just slightly thicker. 

\section{Summary and discussion}
We have presented \texttt{MoCA}, a Monte Carlo code for Comptonisation in Astrophysics which includes polarisation. To our knowledge \texttt{MoCA} is the first code operating with single photons and including all special relativity and quantum effects.
The main disadvantage of this approach is the long computing time, which implies the need to parallelise the code on clusters of computers.
The advantage with respect to pure analytical models such as those available in XSPEC is that we can explore the totality of the parameters space for the Comptonising medium (i.e. thermal energy and optical thickness of the corona) without any restriction and this approach will also allow a better understanding of the whole process.
We also included all corrections such as Klein-Nishina cross-section and scattering angle distribution. 
These effects, small below 100 keV, must nonetheless be taken into account when inferring the thermal energy of the corona from observations. In some sources it has been found that coronae can have extremely high energy cut-off \citep[e.g. NGC 5506,][]{Matt2015} and therefore thermal energy, which is inferred by measuring the curvature of NuSTAR spectra at high energy and in this context K-N effects cannot be neglected. From the polarimetric point of view we did not seen any deviation due to K-N effects, but this was to expected as we focussed our attention below 100 keV where these effects are small. However, one can imagine a scenario in which the thermal energy of the corona is few tens of keV and in that case we expect to see a difference both on the spectrum and the polarisation but we defer such investigation to future papers focussed on the exploration of the coronal parameters space. 
In its actual form the code is fast enough to explore different geometries of the corona with different parameters. Spectra can then be compared with those obtained by NuSTAR to derive coronal parameters, especially in the high optical depth regime where analytical models are not reliable. 
As already mentioned, much observational evidence points in the direction of compact coronae above or around the compact object.
In order to properly treat such coronae, gravitational effects must be taken into account. We have recently included a ray-tracing routine to take into account GR effects: this new version of \texttt{MoCA}, and applications of the code to different astrophysical scenarios, will be discussed in future papers. Nonetheless we have shown the geometrical effect of more compact coronae on the spectra and the polarisation signal: the spectra become softer as the corona shrinks and the polarisation changes dramatically as we approach a more symmetrical shape of the corona. The study we performed will also be useful to quantify the impact of GR effects on compact coronae with respect to a purely geometrical effect.

In a few years, when the NASA Imaging X-ray Polarisation Explorer, IXPE, is launched, it will be possible to use our code to simulate the polarisation signal expected in different scenarios and to compare our results with the observation.
We show that polarimetry has the potential to discriminate between different geometries of the corona shedding a light on the possible origin of the corona itself. Even integrating the polarisation signal in the next generation X-ray polarimeters energy band the two geometries we considered in this work are expected to produce different signals at any inclinations.
We also show that if the photons arising from the disc are initially polarised, the final signal can be very different from the unpolarised seed photons scenario.
It will be fundamental to verify the nature of the polarisation expected from the accretion disc by observing a very bright XRB in a clean soft state before observing the polarisation of Comptonised radiation in hard state.
\texttt{MoCA} is a Comptonisation code which calculate the scattering-induced polarisation produced by such mechanism.
Nonetheless if magnetic fields are present, synchrotron X-ray radiation is expected to be polarised as well. However we show that scattering-induced polarisation is of the order of few \% while synchrotron radiation is expected to be polarised up to several tens \% so the two mechanisms should be clearly distinguishable. 

\begin{appendix}
\section{Description of the code}
In this Appendix we give a detailed description of how the code works when applied to accreting systems by dividing the process in three main steps: 
the production of seed photons from the accretion disc, the parametrisation of the scattering material and the treatment of the scattering.

\subsection{\label{seed}Seed photons}
Up to now, we have implemented two different geometries for the source of the seed photons: either a point-like along the symmetry axis of the system, or an accretion disc.
The point-source have been used only for testing in this work while the disc produces an MTBB
, with the radial dependence of the temperature as for a Shakura-Sunyaev \citep{Shakura1973} geometrically thin and 
optically thick disc:

\begin{eqnarray}
 \label{bbody}
 T(R) &=& \left[ \dfrac{3GM\dot{m}}{8\pi R^3 \sigma_{SB}} \left( 1- \sqrt{\dfrac{R_{in}}{R}}\right) \right]^{\dfrac{1}{4}} 
,\end{eqnarray}
where $G$ is the gravitational constant, $\sigma_{SB}$ the Stefan-Boltzmann constant, $M$ the mass of the central compact object, $\dot{m}$ the accretion rate and $R_{in}$ the inner radius of the disc, often assumed to be coincident with the ISCO (the innermost stable circular orbit).
The first routine involved in the production of seed photons is devoted to extract the radius at which the photon 
is emitted, in order to calculate the temperature at that radius (Eq.~\ref{bbody}), thence the energy of the photon (see below).

In our code we have made extensive use of what we call the fundamental Monte Carlo approach which consists in randomly extracting the continuous variable $x$ according to its probability distribution, $F(x)$, within the interval $\left[x_{min},x_{max}\right]$. This can be expressed by the following formula:
\begin{equation}
\label{MCapproach}
 \xi=P(x)=\dfrac{\int_{x_{min}}^x F(x)\,dx}{\int_{x_{min}}^{x_{max}} F(x)\,dx}
\end{equation}
where $P(x)$ is the probability density function (PDF) of $x$, normalised to one and hence associated with a random number, $\xi$, belonging to the interval $\left[0,1\right]$.
If $F(x)$ is analytically integrable and the integral is invertible, the aleatory variable, $x$, is a function of the random number and it can be easily calculated with an extraction of $\xi$. On the other hand if the integral is not invertible we can still build a grid of value for $P(x) \in \left[0,1\right]$ as a function of $x$ and then extract a random number $\xi (=P(x))$ which can be associated to the corresponding value of $x$ in the grid; we will refer to the latter procedure as the tabular approach.

The number of photons arising from an annulus $R dR$ emitting as a black body 
is proportional to $T(R)^3$ so Eq.~\ref{MCapproach} becomes

\begin{equation}
\label{MCradial}
 \xi=P(R)=\dfrac{\int_{R_{in}}^R T(R)^3\,RdR}{\int_{R_{in}}^{R_{out}} T(R)^3\,RdR}
\end{equation}
where $R_{in}$ and $R_{out}$ are the inner and outer radius of the disc respectively and T(R) is given by Eq.~\ref{bbody}.
Integration of Eq.~\ref{MCradial} gives

\begin{equation}
\label{radial}
 \xi=P(R)=\dfrac{\sqrt{\pi} \, \Gamma \left( \dfrac{7}{4} \right) - B \left( \sqrt{\dfrac{R_{in}}{R}}, \dfrac{1}{2}, \dfrac{7}{4} \right) \, \Gamma \left( \dfrac{9}{4} \right)}{\sqrt{\pi} \, \Gamma \left( \dfrac{7}{4} \right) - B \left( \sqrt{\dfrac{R_{in}}{R_{out}}}, \dfrac{1}{2}, \dfrac{7}{4} \right) \, \Gamma \left( \dfrac{9}{4} \right)}
\end{equation}
where $\Gamma$ is the Gamma function and $B$ is the incomplete beta function.
Once the radius is determined, the energy of the seed photon, $h\nu$, is extracted from the corresponding Planck distribution as a function of the temperature
at that radius (Eq.~\ref{bbody}):
\begin{equation}
 \label{Planck}
 B_{\nu}(T(R)) = \dfrac{2\,h\,\nu^3}{c^2}\dfrac{1}{e^{\frac{h\,\nu}{k_B\,T(R)}}-1}
\end{equation}
where $c$ is the speed of light and $k_B$ the Boltzmann constant.
Integration of Eq.~\ref{Planck} gives the Error function, a non-elementary function defined only in integral form or by Taylor expansion of the integrand.
We used therefore the rejection method (also known as acceptance-rejection sampling) in order to extract the energy of the photon from Eq.~\ref{Planck} once the peak energy at that temperature has been calculated.
After the energy of the seed photon is calculated, the four-wavevector, $K^{\mu} = \left( \dfrac{2\pi E}{hc}, \vec{k} \right)$ and the polarisation four-vector, $P^{\mu} = \left( 0, \vec{p}\right)$ are calculated by another routine. The direction of the emitted photon, $\vec{k}$, is randomly chosen for isotropic emission. The unit four-vector $P^{\mu}$ is only defined to within a multiple of $K^{\mu}$ and its temporal component can always be made equal to zero \citep{Connors1980}.
When interested only in spectra, polarisation calculations can be turned off to save computing time, otherwise $\vec{p}$ is randomly chosen on a plane normal to the direction of the photon, $\vec{k}$, in case of unpolarised radiation or it is chosen according to some initial polarisation but always keeping $\vec{p}$ and $\vec{k}$ perpendicular.
In the case of optically thick atmosphere ($\tau \rightarrow \infty $) the photons are no more isotropically emitted but are instead preferentially emitted perpendicularly to the plane of the disc, the so-called limb darkening effect.
In this case we used the analytical expression given in \citet{Laor1990} which represents a good approximation: 
\begin{equation}
\label{E2.4}
F(\mu)=1+a\,\mu
\end{equation}
where $\mu$ is the cosine of the polar angle, $\theta = \arccos(\mu)$, measured from the axis normal to the disc surface,  $F$ is the flux and $a=2.06$ for an optically thick atmosphere of electrons.
Moreover the optically thick atmosphere of the disc has also the effect of linearly polarise the emitted photon with the polarisation vector parallel to the plane of the disc, with a degree which is $~12\%$ for $\mu$=0 and decreases down to zero for photons emitted perpendicularly to the disc ($\mu = 1$) as in Table XXVI in \cite{Chandrasekhar1960}.

Finally, the Stokes parameters $Q$ and $U$ which describe the degree and angle of linear polarisation: 
\begin{align}
\label{piechi}
\Pi  &= \dfrac{\sqrt{Q^2 + U^2}}{I} \nonumber \\
\chi &=  \dfrac{1}{2} \arctan \dfrac{U}{Q}
\end{align}
can be calculated from the polarisation vector by making use of auxiliary unit vectors following the approach of \citet{Matt1996}.

\subsection{\label{corona}Corona parametrisation}
The corona of electrons surrounding the disc and responsible for the Comptonisation of seed photons is characterised by its thermal energy, $k\,T_e$, its geometry and its density (or optical depth).
We considered two geometries for the corona: a sphere and a very oblate spheroid which we will refer to as a slab (both sketched in Fig.~\ref{geometries}, not to scale).
Instead of using a value for the density of the free electrons composing the corona, $n_e$, we can work with the optical depth $\tau$ whose definition (from e.g. \cite{Rybicki}) is:
\begin{equation}
\label{tau}
d\tau = n_e \, \sigma_{sc} \, dx
\end{equation}
where $\sigma_{sc}$ is the scattering cross-section (Klein-Nishina, in our case) and $dx$ is the space travelled by the photon.
For both the slab and spherical geometry the optical depth, $\tau$ is defined along the vertical of the symmetry axis, as in \texttt{compPS}. This means that a photon travelling parallel to the disc in the slab corona will see a much higher optical thickness while for the spherical corona it will be the same, being radially symmetric by definition.
In our simulations the slab corona has an height of 10 $r_g$ and covers the whole accretion disc up to 1000 $r_g$. This, in turn, means that a photon travelling through the corona parallel to the disc will see an optical depth 100 times larger than if it travels vertically.
Lastly, to determine the velocity of free electrons composing the corona we used the Maxwell-J\"{u}ttner distribution as given in  \cite{Titarchuk1995}
\begin{equation}
\label{MaxJutt}
 f(v)=\dfrac{\gamma^5 \,e^{-\gamma / \Theta}}{4\,\pi \Theta \, K_2(1/\Theta)}  \mbox{,  } \, \Theta=\dfrac{k\,T_e}{m_e\,c^2}
\end{equation}
where $\gamma$ is the Lorentz factor of the electron in $mc^2$ units and $K_2$ is the modified Bessel function of the second type.

\subsection{\label{scattering}Coronal scattering}
The Compton scattering differential cross-section for unpolarised radiation is given by the Klein-Nishina formula:
\begin{equation}
\label{KNunpol}
\dfrac{d\sigma_{KN,U}}{d\Omega} = \dfrac{r_0^2}{2}\dfrac{\epsilon^{\prime 2}_1}{\epsilon^{\prime 2}} \left( \dfrac{\epsilon^{\prime}}{\epsilon^{\prime}_1}+\dfrac{\epsilon^{\prime}_1}{\epsilon^{\prime}}-\sin^2\Theta_{sc} \right) 
\end{equation}
where $r_0$ is the classical radius of the electron, $\epsilon^{\prime}$ and $\epsilon^{\prime}_1$ are the energies of the incident and scattered photons in the reference
frame of the electron\footnote{Here and in the rest of the paper we use the following convention: `primed' quantities are in the reference frame of the electron, while the subscript 1 means `after the scattering'.} and $\Theta_{sc}$ is the scattering angle.
The differential Compton scattering cross-section for a free electron at rest in the case of a 100$\%$ polarised incident beam can be written as
\begin{equation}
\label{KNpol}
\dfrac{d\sigma_{KN,P}}{d\Omega} = \dfrac{r_0^2}{2}\dfrac{\epsilon^{\prime 2}_1}{\epsilon^{\prime 2}}\left( \dfrac{\epsilon^{\prime}}{\epsilon^{\prime}_1}+\dfrac{\epsilon^{\prime}_1}{\epsilon^{\prime}}-2\sin^2\Theta_{sc}\cos^2\Phi_{sc} \right) 
\end{equation}
where $\Phi_{sc}$, the azimuthal scattering angle, is defined as the angle between the polarisation vector of the incident photon and the plane of scattering (defined by the directions of the incident and scattered photons).
The four-wavevector $K^{\mu} = \left( \dfrac{2\pi E}{hc}, \vec{k} \right)$, and the polarisation four-vector $P^{\mu} = \left( 0, \vec{p}\right)$,  are Lorentz transformed in the reference frame of the electron \citep[see][]{Connors1980}.
The first transformation was needed in order to obtain the energy of the incident photon in the reference frame of the electron, $\epsilon^{\prime}$, necessary to calculate the Klein-Nishina cross-section (Eq.~\ref{KNunpol}) and therefore the mean free path (MFP) of the photon.

The MFP is defined as the average distance a photon can travel through an absorbing or scattering material without being absorbed or scattered.
Assuming that the electrons are spatially Poisson (i.e. randomly) distributed, the probability that a scattering happens is
\begin{equation}
\label{probscat}
P=1-e^{-\tau}
\end{equation}
where $\tau$ is the optical depth of the medium defined previously (see Eq.~\ref{tau}).
The inverse of the product between the density and the cross-section of the scattering material is nothing else that the MFP $l= \dfrac{1}{n_e\,\sigma_{sc}}$.
On the other hand, the MFP can also be geometrically defined as the ratio between the size of the scattering medium and its optical depth,
which allows us to work without explicitly using the electron density, $n_e$. 
In the case of the slab corona, the optical depth of the medium was defined with respect to the semi-diameter of the spheroid along the spin-axis while for the spherical corona is defined with respect to the semi-diameter on the azimuthal plane. We can therefore define $l$, respectively, as
\begin{align}
\label{mfp}
l &=  \dfrac{H_{cor}}{\tau}\nonumber \\
l &=  \dfrac{R_{cor}}{\tau}.
\end{align}
Combining the definition of $\tau$ and $l$ in Eq. (\ref{probscat}) and inverting we can derive the space, $x$, travelled by the photon before the scattering happens:
\begin{equation}
\label{space}
x = - \ln(1-P) \, l
.\end{equation} 

Finally, the photon energy dependence on the MFP is taken into account by weighting the travelled space with the ratio between the total Thomson cross-section 
and the total Klein-Nishina cross-section, so that the MFP increases with energy because $\sigma_{KN}$ decreases. 
Therefore, the travelled space, $x$, is calculated as 
\begin{equation}
\label{finspace}
x = - \ln(1-\xi) \, l \, \dfrac{\sigma_T}{\sigma_{KN}}
\end{equation} 
where $\xi$ is a random number between $0$ and $1$, $l$ is calculated from Eq.~\ref{mfp} and $\sigma_{KN}$ is the total Klein-Nishina cross-section obtained by 
integration of Eq~\ref{KNunpol}, explicitly
\begin{eqnarray}
\label{sigmaKN}
\sigma_{KN} &=& \sigma_T \, \dfrac{3}{4} \left\{ \dfrac{1+y}{y^3} \left[ \dfrac{2y\,(1+y)}{1+2y} - ln(1+2y) \right] \right. \nonumber  \\
 &+& \left.\dfrac{1}{2y} \ln(1+2y) - \dfrac{1+3y}{(1+2y)^2}\right\} 
\end{eqnarray}
where $y=\dfrac{\epsilon^{\prime}}{mc^2}$ is the ratio between the energy of the incident photon in the reference frame of the electron
and the rest mass energy of the electron.

The new coordinates of the photon were then calculated. If the arrival point falls within $R_{in}$, the photon is considered lost inside the black hole. 
If the arrival point is on the disc, the photon is absorbed and considered lost as well 
(no reflection from the accretion disc is, at moment, included). 
If the arrival point falls outside the corona, the photon escapes towards the observer with no scattering. The Stokes parameters Q and U were calculated and saved together with the energy and direction of the photon.
Lastly, if the arrival point falls within the corona, the scattering happens.

\begin{figure}
\includegraphics[width=88.mm,clip=]{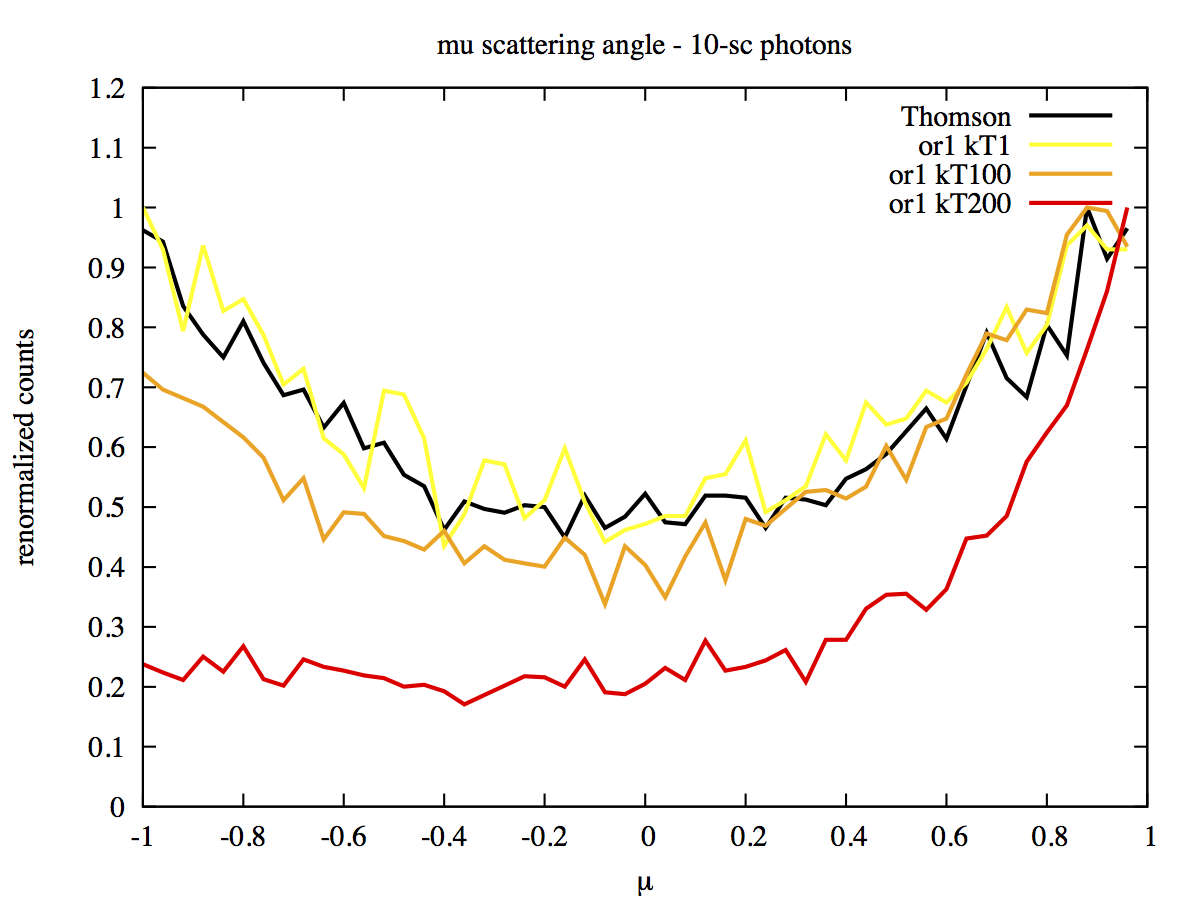}
\caption{\label{plot_musc} $\mu = \cos(\Theta_{sc})$ distribution for our test case (see text for more details) applied to photons which experienced already 10 scattering events. Black line correspond to Thomson formula while coloured lines correspond to K-N formula. In Thomson regime forward and backward scattering are equally probable and twice as likely to happen than a 90 degree scattering. In K-N regime, as the energy of the electrons increases with respect to the energy of the photon, forward scattering is favoured with respect to backward scattering (red line).}
\end{figure}

In the latter case, the first quantity to be calculated is the polar scattering angle $\Theta_{sc}$.
$\Theta_{sc}$ is independent of the polarisation status of the incident photon and it is obtained by Eq. ~\ref{KNunpol} following \cite{Matt1996}.
The probability, $P$, for a photon to scatter at an angle, $\Theta_{sc}$ is
\begin{eqnarray}
\label{thetasc_KN}
P&=& \dfrac{1}{\dfrac{2}{3}+2y+\dfrac{1}{y}\ln\left(1+2y\right)} \left[ y\left(\dfrac{3}{2}+\mu-\dfrac{1}{2}\mu^2\right) + \right. \nonumber \\
&+& \dfrac{1}{3}\left(1+\mu^3\right)-\dfrac{1}{y}\{\ln\left[1+y\left(1-\mu\right)\right] + \nonumber \\
&-& \left. \ln\left(1+2y\right) \}\right]
\end{eqnarray} 
where $\mu = \cos(\Theta_{sc})$.
$P$ can always be associated to a random number $\xi$ belonging to the interval $\left[0,1\right]$, as usual.
Eq. ~\ref{thetasc_KN} cannot be analytically inverted so we follow the tabular approach by building a grid of values for $P$ as a function of $\mu$ and then, by extracting a random number between 0 and 1 we get the corresponding cosine of the scattering polar angle.
On the other hand, when we use Thomson approximation the polar scattering angle is calculated integrating the well-known Thomson scattering cross-section:
\begin{equation}
\label{thomson}
\dfrac{d\sigma_{T}}{d\Omega} = \dfrac{q^2}{4 \pi \epsilon^{\prime} mc^2} \dfrac{1 + \cos^2(\Theta_{sc})}{2}
\end{equation}
where $q$ and $m$ are the charge and mass of the electron and  $\epsilon^{\prime}$ the energy of the photon in the reference frame of the electron.
Integration of Eq.~\ref{thomson} over $d\Omega$ gives the probability, $P$, for a photon to scatter at an angle, $\Theta_{sc}$:
\begin{equation}
\label{thetasc_Thom}
P = \dfrac{1}{8}\left( \mu^3 + 3\mu + 4 \right)
\end{equation}
We performed some tests using a point source at the origin (no compact object present) emitting monochromatic photons at 1 keV and surrounded by a slab corona with $H_{cor}/R_{cor} = 0.01$. In Fig.~\ref{plot_musc}  we show the distribution of $\cos(\Theta_{sc})$ for the subsample of photons which had already experienced ten scattering events. The black line shows the distribution in Thomson regime (Eq.~\ref{thetasc_Thom}) which gives equal probability of having forward ($\mu = 1$) or backward  ($\mu = -1$) scattering, this probability being about twice the probability of having a 90 degrees scattering angle (i.e. $\mu = 0$).
In coloured lines the same distribution in K-N regime (Eq.~\ref{thetasc_KN}) for increasing energy of the electron as indicated in the figure.
When the energy of the electron is comparable with the energy of the photon (yellow line) we are naturally in Thomson regime and  Eq.~\ref{thetasc_KN} reduces to Eq.~\ref{thetasc_Thom} giving the same distribution. As the energy of the electrons increases, forward scattering is favoured with respect to backward scattering (red line), as expected.

\begin{figure}
\includegraphics[width=88.mm,clip=]{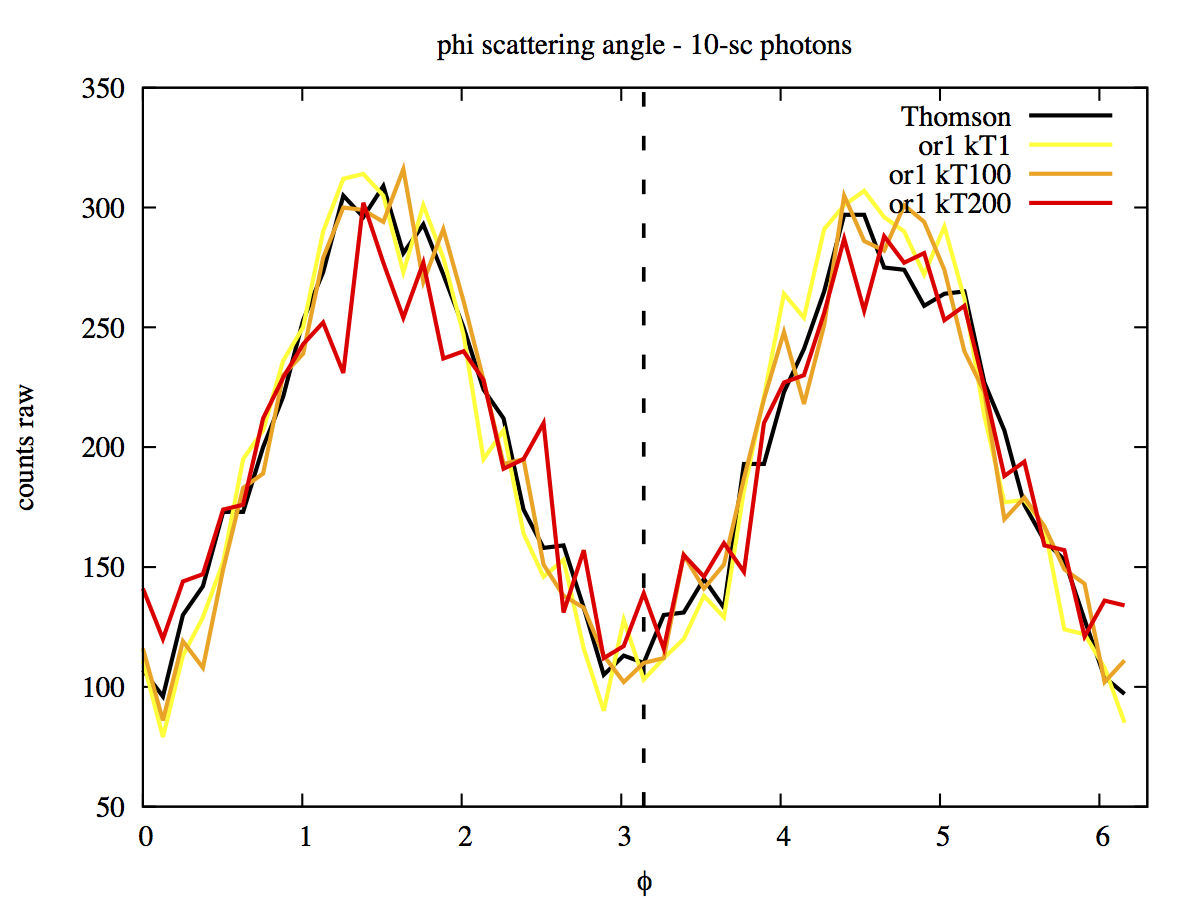}
\caption{\label{plot_phisc} $\Phi_{sc}$ distribution for our test case similarly to Fig.~\ref{plot_musc}. The azimuthal scattering angle distribution does not depend on the energy but it is dramatically reduce as $\Theta_{sc}$ tends to zero. We did, indeed, notice a slight reduction of the modulation for the photons which are more likely forward scattered (red line). However this effect is is significant only for photons scattered at  $\Theta_{sc} \sim 0$.}
\end{figure}

Once $\Theta_{sc}$ is calculated, the energy exchange can be easily derived by the Compton formula as
\begin{equation}
\label{compton}
\epsilon^{\prime}_1 = \dfrac{\epsilon^{\prime}}{1+\dfrac{\epsilon^{\prime}}{mc^2}(1-\cos\Theta_{sc})}
\end{equation}
where $\epsilon^{\prime}$ and $\epsilon^{\prime}_1$ are, respectively, the energies of the incident and scattered photon in the reference frame of the electron. 
Then, once $\Theta_{sc}$ and $\epsilon^{\prime}_1$ are known, we use Eq.~\ref{KNpol} to derive the azimuthal scattering angle as in \cite{Matt1996}.
\begin{equation}
\label{phisc}
\left(2\pi P-\Phi_{sc}\right)\left(\dfrac{\epsilon^{\prime}_1}{\epsilon^{\prime}}+\dfrac{\epsilon^{\prime}}{\epsilon^{\prime}_1}-\sin^2\Theta_{sc}\right)+\sin^2\Theta_{sc}\sin\Phi_{sc}\cos\Phi_{sc}=0
\end{equation} 
We note that in a single-photon approach each photon is polarised. If the radiation is overall unpolarised, it simply means that photons will have random polarisation which will give a zero net polarisation \citep{Angel1969}.
Eq.~\ref{phisc} describes the azimuthal modulation. 
This equation cannot be analytically inverted and we use the same tabular approach to obtain the azimuthal scattering angle.
In Fig.~\ref{plot_phisc} we show the azimuthal scattering angle distribution (Eq.~\ref{phisc}) similarly to Fig.~\ref{plot_musc}.
The modulation of $\Phi_{sc}$ does not depend on the relative energy between the photon and the electron \citep[see e.g. Fig. 3 in][]{Matt1996} but it is strongly affected by $\Theta_{sc}$. When this is zero the modulation decreases \citep[Fig. 1 in][]{Matt1996} and we slightly see this effect by looking at the red line in Fig.~\ref{plot_phisc} which belongs to the photons with a higher probability of being scattered forwards. However, this effect is significant only for photons scattered at $\Theta_{sc} \sim 0$.

Once the scattering angles are calculated, by making use of auxiliary vectors as described in \cite{Matt1996}, 
the direction of the scattered photon can be calculated in the reference frame of the electron.
The scattering angles, together with the energy of the scattered photon, allow us also to calculate the degree of polarisation induced by Compton scattering:
\begin{equation}
\label{pie}
\Pi = 2\dfrac{1-\sin^2\Theta_{sc}\cos^2\Phi_{sc}}{\dfrac{\epsilon^{\prime}_1}{\epsilon^{\prime}} + \dfrac{\epsilon^{\prime}}{\epsilon^{\prime}_1} - 2\sin^2\Theta_{sc}\cos^2\Phi_{sc}}
.\end{equation}
If the value of a random number is greater than $\Pi$, the scattered photon's polarisation vector, $\vec{p^{\prime}_1}$,
is randomly chosen on a plane perpendicular to its direction, otherwise the polarisation vector  
is calculated using \cite{Angel1969}:
\begin{equation}
\label{angel}
\vec{p^{\prime}_1} = \dfrac{1}{|\vec{p^{\prime}_1}|} \left[ \left( \vec{p^{\prime}} \times \vec{k^{\prime}_1} \right) \times \vec{k^{\prime}_1} \right] 
\end{equation}
where $\vec{k^{\prime}_1}$ is the spatial part of the scattered photon's four-wavevector in the reference frame of the electron.
Eventually, the four-wavevector and the polarisation vector of the photon are anti-transformed to the reference frame of the disc and the photon
keeps travelling until the next scattering happens or until it falls into the compact object, onto the disc or until it reaches the observer at infinity.
For each photon which reaches the observer at infinity the Stokes parameters Q and U are calculated and registered together with the direction, energy and number of scatterings experienced by the photon before escaping.

\end{appendix}

\section{Acknowledgements}
We would like to thank Rene Goosmann for very useful and stimulating discussions. We acknowledge financial support from the European
Union Seventh Framework Programme (FP7/2007-2013) under grant agreement n.312789. FT would like to thank the anonymous referee for the constructive discussions which significantly improved the quality of the paper.

\bibliographystyle{aa}
\bibliography{newbib} 
\end{document}